\documentclass[aps,prd,amsmath,amssymb,showpacs]{revtex4} 
\usepackage{graphicx}
\usepackage{epsfig,float}
\usepackage{dcolumn}
\usepackage{bm}
\usepackage{bbm}
\usepackage{morefloats}
\usepackage{color}
\usepackage{amsmath}
\usepackage{mathrsfs}
\usepackage{slashed}
\usepackage{multirow}
\usepackage{amssymb}
\usepackage{bbding}
\usepackage{hyperref}
\usepackage{appendix}
\usepackage{multirow}
\usepackage{graphicx}
\usepackage{caption}
\usepackage{subfigure}
\usepackage{subfloat}    
\usepackage{float}
\usepackage[font=footnotesize]{caption}

\graphicspath{{./image}}

\makeatletter
\newcommand*{\rom}[1]{\expandafter\@slowromancap\romannumeral #1@}
\makeatother

\begin{document}
  
\title{ Study of weak radiative decays of $D^0\to V \gamma$}

\author{Ye Cao$^{1,2}$\footnote{Email: caoye@ihep.ac.cn} and Qiang Zhao$^{1,2}$\footnote{E-mail: zhaoq@ihep.ac.cn} }

\affiliation{ 1) Institute of High Energy Physics,
        Chinese Academy of Sciences, Beijing 100049, P.R. China}

\affiliation{ 2) University of Chinese Academy of Sciences, Beijing 100049, P.R. China}

\begin{abstract}

The weak radiative decay of $D^0\to V\gamma$ with $V=\bar{K}^0$, and $\phi$, $\rho^0$, and $\omega$, is systematically studied in the vector meson dominance (VMD) model. It allows us to distinguish the short-distance mechanisms which can be described by the tree-level transitions in the non-relativistic constituent quark model, and the long-distance mechanisms which are related to the final-state interactions (FSIs). We find that the FSI effects play a crucial role in $D^0\to V\gamma$ and the SU(3) flavor symmetry can provide a natural constraint on the relative phase between the short and long-distance transition amplitudes. Our analysis suggests that the $D$ meson weak radiative decays can serve as a good case for investigating the non-perturbative QCD mechanisms at the charm quark mass region.
 
\end{abstract}

\maketitle

\section{introduction}

The charm quark production and decay has been an ideal place for probing the non-perturbative QCD effects due to the reason that the charm quark mass is not heavy enough. During the past years the accumulation of large $D$ meson samples allows more precise measurements of its exclusive decays and also access more channels which have not been measured before. One of those interesting processes is the weak radiative decay of $D^0\to V\gamma$ with $V$ denoting the light vector mesons $\bar{K}^{*0}$, and $\phi$, $\rho^0$, $\omega$. These four exclusive decays are either Cabibbo-favored ($D^0\to \bar{K}^{*0}\gamma$) or singly Cabibbo-suppressed ($D^0\to \rho^0\gamma/\omega\gamma/\phi\gamma$). 

It is interesting to note that the $D$ meson weak decays would be very different from $B$. The $B$ meson weak radiative decays has been studied in the framework of an effective Hamiltonian approach~\cite{Grinstein:1987vj,Grinstein:1990tj}, where the approximation of free quark transition, namely the transition amplitude is dominated by the approximately free quark decay, e.g. $b\to s\gamma$, is well confirmed. In contrast, the dominance of the free quark transition picture cannot be justified for $D\to V\gamma$ due to the relatively light charm quark mass. As a natural consequence, non-perturbative contributions may become important in $D\to V\gamma$ and a direct confirmation of such effects should provide crucial informations about the decay mechanisms.

In experiment, the decay of $D^0\to\omega\gamma$ was first searched by CLEO-c in Ref.~\cite{CLEO:1998mtp} but with only an upper limit set, i.e. $ BR(D^0\to\omega\gamma)<2.4\times 10^{-4}$. The two decay channels, $\phi\gamma$ and $\bar{K}^{*0}\gamma$, were measured by the BaBar Collaboration in 2008~\cite{BaBar:2008kjd} with $BR(D^0\to\phi\gamma)=(2.78\pm 0.30\pm 0.27)\times 10^{-5}$ and $BR(D^0\to\bar{K}^{*0}\gamma)=(3.28\pm 0.20\pm 0.27)\times 10^{-4}$. The Belle Collaboration confirmed the BaBar result for $D^0\to \phi\gamma$~\cite{Belle:2016mtj}. However, the branching ratio of  $D^0\to \bar{K}^{*0}\gamma$ measured by the Belle Collaboration, i.e. $BR(D^0\to\bar{K}^{*0}\gamma)=(4.66\pm 0.21\pm 0.21)\times 10^{-4}$, turns out to be significantly different from that from BaBar. Apart from the  $\phi\gamma$ and $\bar{K}^{*0}\gamma$ channels, Belle also measured the decay of $D^0\to\rho^0\gamma$, i.e. $BR(D^0\to\rho^0\gamma)=(1.77\pm 0.30 \pm 0.07)\times 10^{-5}$~\cite{Belle:2016mtj}, which is the same order of magnitude as $D^0\to \phi\gamma$. This may be reasonable since both processes are color-suppressed and  singly Cabibbo-suppressed.

Theoretical studies of the $D^0\to V\gamma$ can be found in the literature. In Ref.~\cite{Asthana:1990kr} a modified quark model was applied to estimate the decay of $D^0\to\bar{K}^{*0}\gamma$ which had underestimated the data by about a factor of 5. In Ref.~\cite{Cheng:1994kp} an effective Lagrangian approach was developed for dealing with the weak radiative decays of heavy flavor hadron involving the bottom quark. Its extension to $D^0\to\bar{K}^{*0}\gamma$ also yields an underestimated result by about a factor of 4. In Ref.~\cite{Bajc:1994ui} the decay of $D^0\to\bar{K}^{*0}\gamma$ was studied by combining the heavy quark effective field theory and chiral Lagrangian approach. Although the theoretical prediction was consistent with the later measurement, it had large uncertainties which covered a range larger than the experimental value with errors. In Ref.~\cite{Burdman:1995te} a detailed analysis of the short-distance contributions to $D^0\to V\gamma$ via the free $c\to u\gamma$ transition was presented. It was also discussed that the long-distance contributions via the pole terms and VMD should be non-negligible. Similar approaches based on the heavy quark effective theory and chiral Lagrangians was presented in Refs.~\cite{Fajfer:1998dv,deBoer:2017que,Biswas:2017eyn}, where the long-distance contributions are through the pole terms and vector meson dominance (VMD). It is interesting to see that in all these approaches the predicted results are systematically lower than the experimental data by a couple of times up to nearly one order of magnitude. In Ref.~\cite{Shen:2013oua} a covariant light cone approach was adopted for calculating the $D\to V, \ A, \ T$ form factors among  which the $D\to V$ transition can contribute to the weak radiative decay of $D\to V\gamma$. Again, it shows that the the long-distance contributions from the VMD are dominant over the short-distance $c\to u\gamma$ dynamics. Meanwhile, the VMD contributions are still insufficient for accounting for the experimental data.

In this work we are motivated to make a combined analysis of the Cabibbo-favored and singly Cabibbo-suppressed decays of $D\to V\gamma$. Different from other approaches in the literature~\cite{Fajfer:1998dv,deBoer:2017que,Biswas:2017eyn}, where the heavy quark effective Lagrangians are employed to describe the weak couplings for $D^0\to VV$, we calculate the couplings of the flavor-neutral vector meson decays of $D^0\to VV$ in the non-relativistic constituent quark model (NRCQM). It means that the breaking of the SU(3) flavor symmetry will arise partially from the quark model wave functions due to the quark mass difference. Despite the tree-level contributions, we argue that the large b.r.s of some of those intermediate hadronic two-body decays, e.g. $D\to VV$, $VP$, and $PP$, etc., imply that the final-state interactions (FSIs) via rescatterings should be important. Note that the mass threshold of $K^{*+}K^{*-}$, which involves the direct emission of $K^{*+}$ and is sizeable, is almost degenerate with that of $\phi\rho^0$ and $\phi\omega$. As being investigated recently in Ref.~\cite{Cao:2023csx}, the FSIs play a crucial role in the understanding of the puzzling polarization results in $D\to VV$. We will show in this work that this mechanism also plays a crucial role in the description of $D^0\to V\gamma$ and provides a natural source for filling the deficit between the theoretical calculations~\cite{Fajfer:1998dv,deBoer:2017que,Biswas:2017eyn} and experimental measurements~\cite{CLEO:1998mtp,BaBar:2008kjd,Belle:2016mtj}.

As follows, we first introduce the formalism in Sec. II. The numerical results and discussions are presented in Sec. III, and a brief summary is given in the end.

\section{FORMALISM}

The most general form of the $S$-matrix element for a generic radiative weak decay of kind $D(p)\to V(k_1,\ \epsilon_1)+\gamma(k_2,\ \epsilon_2)$, consistent with gauge invariance, is 
\begin{align}
    \begin{split}
        S=I-i(2\pi)^4\delta^4(p-k_1-k_2)(\mathcal{M}^{\text{(PC)}}+\mathcal{M}^{\text{(PV)}}),
        \label{Eq: S-matrix}
    \end{split}
\end{align}
where
\begin{align}
    \mathcal{M}^{\text{(PC)}}=i {\mathcal A}^{\text{(PC)}}\epsilon_{\alpha\beta\mu\nu}k_1^{\alpha}p^{\beta}\varepsilon_1^{*\mu}\varepsilon_2^{*\nu},
    \label{Eq: F^PC}
\end{align}
and 
\begin{align}
    \mathcal{M}^{\text{(PV)}}={\mathcal A}^{\text{(PV)}}(p_{\mu}p_{\nu}-g_{\mu\nu}k_2\cdot p)\varepsilon_1^{*\mu}\varepsilon_2^{*\nu},
    \label{Eq: F^PV}
\end{align}
where $p$, $k_1$, and $k_2$ are the four-momenta of the initial and the final mesons and the photon, respectively. $\varepsilon_1^{\mu}$ and $\varepsilon_2^{\nu}$ are the vector-meson and the photon polarization vectors. $\mathcal{M}^{(\text{PC})}$ and $\mathcal{M}^{(\text{PV})}$ are the parity-conserving (PC) and parity-violating (PV) amplitudes. The parity-conserving amplitude involves $P$ wave in the final state
while the parity-violating amplitude involves $S$ and $D$ waves. From Eq. (\ref{Eq: S-matrix})-(\ref{Eq: F^PV}), the decay rate is calculated to be 
\begin{align}
    \begin{split}
        \Gamma(D\to V\gamma)=\frac{|\boldsymbol{k}_2|^3}{4\pi}(|\mathcal{A}^{\text{(PC)}}|^2+|\mathcal{A}^{\text{(PV)}}|^2),
    \end{split}
\end{align}
where $|\boldsymbol{k}_2|={(m_D^2-m_V^2)}/{2m_D}$ is the decay momentum in the rest frame of $D$ meson.

\begin{figure}
    \centering
    \subfigure[]{
    \includegraphics[width=4.5cm]{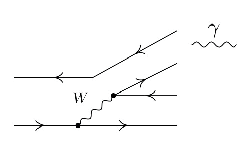}}
    \subfigure[]{
    \includegraphics[width=3cm]{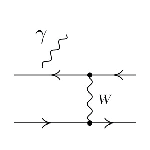}}
    \subfigure[]{
    \includegraphics[width=3cm]{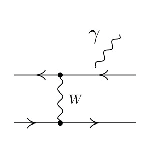}}
    \subfigure[]{
    \includegraphics[width=3cm]{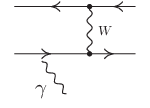}}
    \subfigure[]{
    \includegraphics[width=3cm]{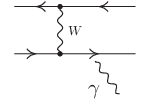}}
    \caption{Schematic diagrams for the process $D^0\to V\gamma$ at the quark level.}
    \label{fig: QuarkLevel diagrams}
\end{figure}

\begin{figure}
    \centering
    \subfigure[]{
    \includegraphics[width=4.2cm]{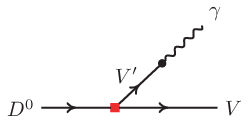}}
    \subfigure[]{
    \includegraphics[width=4.2cm]{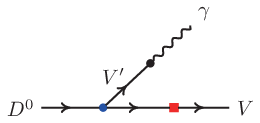}}
    \subfigure[]{
    \includegraphics[width=4.2cm]{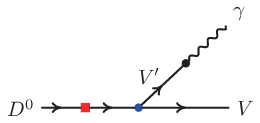}}
    \subfigure[]{
    \includegraphics[width=4.5cm]{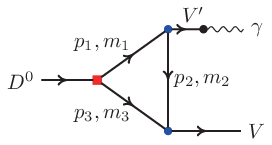}}
    \caption{Schematic diagrams for the process $D^0\to V\gamma$ in the VMD model at the hadronic level. (a) stands for the tree-level transitions; (b) and (d) denote two types of pole-term contributions: type-I (b) and type-II (c); (d) stands for the hadronic triangle loop transitions. Red squares, blue dots, and black dots represent weak, strong, and electromagnetic vertices respectively.}
    \label{fig: D-V gamma}
\end{figure}

At the tree level there are two types of two-body radiative decay diagrams  at the quark level as illustrated in Fig. \ref{fig: QuarkLevel diagrams}. The first corresponds to the internal $W$-emission process $c\to q_1\bar{q}_2q$, followed by $\bar{q}_2q\to\gamma$ (or $q_1\bar{u}\to\gamma$) which is depicted in Fig. \ref{fig: QuarkLevel diagrams}(a). The second one corresponds to the internal conversion processes $c\bar{u}\to q_1\bar{q}_2$ with a photon attached to any of the four quark (antiquark) lines which are depicted in the remaining diagrams, Fig. \ref{fig: QuarkLevel diagrams}(b)-(e). At the hadronic level, these four processes give rise to the set of contributions as shown in Fig.~\ref{fig: D-V gamma} (a)-(c). These are actually the VMD contributions and pole terms considered in some of those previous studies~\cite{Burdman:1995te,Fajfer:1998dv,deBoer:2017que,Biswas:2017eyn}. But as discussed earlier, these contributions are systematically smaller than the experimental data and insufficient for accounting for the observed results. 

A possible solution is the FSIs via the intermediate meson scatterings taking into account the large b.r.s for some of the hadronic two-body decays which can rescatter into $VV$. Then, in the framework of the VMD model, contributions from such a mechanism to $V\gamma$ can be included. It should be stressed that although the FSI contributions are loop corrections, their effects may not be small given that the $D^0\to VV$ couplings are large. To be more specific and practical, we identify the Cabibbo-favored decay channels or those processes involving the DE transitions as the leading intermediate processes which can contribute to $V\gamma$ via the triangle diagram of Fig.~\ref{fig: D-V gamma} (d). 

As follows, we first extract the tree-level amplitudes which contain the contributions from the processes of Fig.~\ref{fig: D-V gamma} (a)-(c), and then extract the amplitudes of the FSIs via the triangle diagram of Fig.~\ref{fig: D-V gamma} (d).

\subsection{Tree-level amplitudes in the VMD model}\label{subsect-tree}

Note that the decay of $D^0\to V\gamma$ only involves the color-suppressed process. The tree-level amplitudes include two parts. One is the internal $W$-emission process (CS-process) (Fig.~\ref{fig: D-V gamma} (a)), and the other ones are the internal conversion via the $W$ exchange, where the pole contributions can be identified (Fig.~\ref{fig: D-V gamma}(b) and (c)).

\subsubsection{CS-processes}

We describe briefly the formalism for extracting the amplitude for the internal $W$-emission process (CS-process) of $D^0\to VV^{\prime}$, which has been constructed in Ref.~\cite{Cao:2023csx}. Then, the flavor neutral vector meson  (we note it as $V^{\prime}$) will propagate as a virtual particle and convert into a photon in the VMD model. The CS-process is essentially the $1\to 3$ flavor-changing emission process at the quark level, accompanied by the color suppression of quark-anti-quark hadronization into final mesons. The corresponding effective weak Hamiltonian operators $\hat{H}_{W,1\to 3}^{(\text{P})}$ can reduce to the form of four-fermion interaction in the non-relativistic approximation, and the detailed expressions which take different forms for parity-conserving (PC) or parity-violating (PV) transitions have been derived in Ref.~\cite{Niu:2020gjw}. The corresponding weak coupling strengths $g_{W(\text{CS})}^{(\text{P})}$ are defined by the transition matrix element as follows:
\begin{align}
    \begin{split}
        i\mathcal{M}^{\text{(P)}}_{\text{CS}}&=\langle V_1(\mathbf{P}_1;J_1,J_{1z})V_2(\mathbf{P}_2;J_2,J_{2z})|\hat{H}_{W,1\to3}^{(\text{P})}|D^0(\mathbf{P}_D;J_i,J_{iz})\rangle\\
        &\equiv g_{W(\text{CS})}^{\text{(P)}}V_{cq}V_{uq},
    \end{split}
\end{align}
where $\mathbf{P}_1=\mathbf{p}_1+\mathbf{p}_4$ and $\mathbf{P}_2=\mathbf{p}_2^{\prime}+\mathbf{p}_3$. The above formula contains spatial wavefunction integrals for which
the NRCQM wavefunctions \cite{Kokoski:1985is,Godfrey:1985xj,Godfrey:1986wj} are adopted. It should be noted that the mass differences within those $VV$ channels will lead to different values for the CS couplings $g_{W(\text{CS})}^{(\text{P})}$ after taking the wavefunction convolutions as a source of
the SU(3) flavor symmetry breaking. In Tab.~\ref{tab: CS waek couplings} the CS couplings $g_{W(\text{CS})}^{(\text{P})}$ of the CF ($D^0\to\bar{K}^{*0}\rho^0/\bar{K}^{*0}\omega$) and SCS ($D^0\to \phi\rho^0/\phi\omega$ and $\rho^0\rho^0/\omega\omega$) decay channels calculated in the NRCQM are listed.

\begin{table}
    \centering
    \caption{The weak couplings $g_{W(\text{CS})}^{(\text{P})}$ in units of $10^{-6}$  for the decays of $D^0\to VV$ including the PC and PV transitions, which is estimated by calculating the CS-process in the NRCQM and the uncertainty comes from the the model parameters $(10\%)$.}
    \label{tab: CS waek couplings}
    \begin{tabular}{lcccccc}
        \hline\hline
        Decay channels             &$D^0\to\bar{K}^{*0}\rho^0/\bar{K}^{*0}\rho^0$              &$D^0\to\phi\rho^0/\phi\omega$             &$D^0\to\rho^0\rho^0/\omega\omega$    \\
        \hline
        $g_{W\text{(CS)}}^{\text{(PC)}}$ [$\text{GeV}^{-1}$]   &$1.47\pm0.31$                                              &$1.63\pm0.34$                             &$1.31\pm0.35$                           \\
        $g_{W\text{(CS)}}^{\text{(PV)}}$ [$\text{GeV}$]        &$1.71\pm0.29$                                              &$1.90\pm0.32$                             &$1.53\pm0.33$                           \\
        \hline\hline
    \end{tabular}
\end{table}

The amplitude of the tree diagram for the parity-conserving (PC) and  parity-violating (PV) transitions shown in Fig. \ref{fig: D-V gamma}(a) are defined as
    \begin{align}
        \begin{split}
            i\mathcal{M}^{(\text{PC})}_{T(a)}&=ig_{DV\gamma}^{(\text{PC})}\epsilon_{\alpha\beta\delta\lambda}p_{\gamma}^{\alpha}p_V^{\beta}\varepsilon_{\gamma}^{\delta}\varepsilon_{V}^{\lambda},\\
            i\mathcal{M}^{(\text{PV})}_{T(a)}&=-ig_{DV\gamma}^{(\text{PV})}\varepsilon_{\gamma}^{\mu}\varepsilon_{V\mu},
        \end{split}
    \end{align}
where the tree-level effective coupling $g_{DV\gamma}^{(\text{P})}$ can be expressed as
\begin{align}
    \begin{split}
        g_{DV\gamma}^{(\text{P})}=-ig_{DVV^{\prime}}^{(\text{P})}\frac{em_{V^{\prime}}^2}{f_{V^{\prime}}}G_{V^{\prime}},
        \label{Eq: gDVgamma}
    \end{split}
\end{align}
where (P) in the above equation can be either (PC) or (PV), $f_{V^{\prime}}$ is the decay constant of vector meson $V^{\prime}$, and it can be extracted using the data for $V^{\prime}\to e^+e^-$~\cite{ParticleDataGroup:2022pth}. We collect the values for $e/f_{V'}$ in Tab.~\ref{table: vector decay constants} for convenience. In the above equation $G_{V^{\prime}}$ is the propagator of the intermediate vector meson $V^{\prime}$,
\begin{align}
    \begin{split}
        G_{V^{\prime}}\equiv \frac{-i}{p_{\gamma}^2-m_{V^{\prime}}^2+im_{V^{\prime}}\Gamma_{V^{\prime}}}=\frac{-i}{-m_{V^{\prime}}^2+im_{V^{\prime}}\Gamma_{V^{\prime}}}.
    \end{split}
\end{align}

\begin{table}
    \centering
    \caption{Vector meson decay constants determined by $V'\to e^+e^-$. The data are taken from the PDG \cite{ParticleDataGroup:2022pth}.}
    \label{table: vector decay constants}
    \begin{tabular}{cccc}
        \hline\hline
        Channel             &Total width of $V'$  &BR($V'\to e^+e^-$)                &$e/f_{V'}(\times 10^{-2})$\\
        \hline
        $\phi\to e^+e^-$    &$4.25$ MeV           &$(2.98\pm 0.03)\times 10^{-4}$   &$-2.26$\\
        $\rho^0\to e^+e^-$  &$147.4$ MeV          &$(4.72\pm 0.05)\times 10^{-5}$   &$6.07$\\
        $\omega\to e^+e^-$  &$8.68$ MeV           &$(7.38\pm 0.22)\times 10^{-5}$   &$1.83$\\
        $J/\psi\to e^+e^-$  &$92.6$ keV           &$(5.97\pm 0.032)\%$   &$2.71$\\
        \hline\hline
    \end{tabular}
\end{table}

\begin{table}
    \centering
    \caption{The tree-level amplitudes of all the Cabibbo-favored and singly Cabibbo-suppressed radiative weak decay channels for $D^0\to V\gamma$ ($V=\bar{K}^{*0},\ \phi,\ \rho^0,\ \omega$). (Note that: there is no CS-process in $D^0\to \rho^0\omega$.)}
    \label{Table: tree-level amplitude}
    \begin{tabular}{cc}
        \hline\hline
        Modes                 &Tree amplitudes (Fig. \ref{fig: D-V gamma}(a))\\
        \hline
        $\bar{K}^{*0}\gamma$  &$\frac{1}{\sqrt{2}}g_{W(\text{CS})}^{\text{(P)}}V_{cs}V_{ud}\frac{em_{\rho^0}^2}{f_{\rho^0}}G_{\rho^0}+\frac{1}{\sqrt{2}}g_{W(\text{CS})}^{\text{(P)}}V_{cs}V_{ud}\frac{em_{\omega}^2}{f_{\omega}}G_{\omega}$ \\
        $\phi\gamma$          &$\frac{1}{\sqrt{2}}g_{W(\text{CS})}^{\text{(P)}}V_{cs}V_{us}\frac{em_{\rho^0}^2}{f_{\rho^0}}G_{\rho^0}+\frac{1}{\sqrt{2}}g_{W(\text{CS})}^{\text{(P)}}V_{cs}V_{us}\frac{em_{\omega}^2}{f_{\omega}}G_{\omega}$\\
        $\rho^0\gamma$        &$-\frac12g_{W(\text{CS})}^{\text{(P)}}V_{cd}V_{ud}\frac{em_{\rho^0}^2}{f_{\rho^0}}G_{\rho^0}+\frac{1}{\sqrt{2}}g_{W(\text{CS})}^{\text{(P)}}V_{cs}V_{us}\frac{em_{\phi}^2}{f_{\phi}}G_{\phi}$\\
        $\omega\gamma$        &$\frac12g_{W(\text{CS})}^{\text{(P)}}V_{cd}V_{ud}\frac{em_{\omega}^2}{f_{\omega}}G_{\omega}+\frac{1}{\sqrt{2}}g_{W(\text{CS})}^{\text{(P)}}V_{cs}V_{us}\frac{em_{\phi}^2}{f_{\phi}}G_{\phi}$\\
        \hline\hline        
    \end{tabular}
\end{table}

The tree-level amplitudes for $D^0\to V\gamma$ ($V=\bar{K}^{*0},\ \phi,\ \rho^0,\ \omega$) can be parametrized in the VMD model as shown in Tab.~\ref{Table: tree-level amplitude}. Note that the vector meson decay constant $f_{V^{\prime}}$ in Tab.~\ref{Table: tree-level amplitude} has a negative sign for $\phi$, and is positive for $\rho^0$ and $\omega$ which can be explicitly determined in the quark model. It suggests that the tree-level amplitude of $D^0\to \omega\gamma$ will be relatively suppressed due to the cancellation between the two terms in Tab.~\ref{Table: tree-level amplitude}. In contrast, the two terms in Tab.~\ref{Table: tree-level amplitude} for $D^0\to\rho^0 \gamma$ will have a constructive interference.

\subsubsection{Pole terms}

The pole contributions to charm meson radiative weak decays are shown in Fig.~\ref{fig: D-V gamma}(b) and (c). At the quark level, they correspond to the internal conversion diagrams $c\bar{u}\to q_1\bar{q}_2$ with a photon attached to any of the four quark lines illustrated by Fig.~\ref{fig: QuarkLevel diagrams}(b)-(e). At the hadronic level, these give rise to the contributions which are part of the long-distance mechanisms. The transition amplitudes are denoted as type-I and type-II, respectively, depending on that the photon emission occurs either before or after the weak transition. It should be stressed that in both cases, all possible spin-one ($J^P=1^\pm$) and spin-zero ($J^P=0^\pm$) intermediate virtual particles can contribute, respectively. Because of this, a reliable calculation of the pole terms is non-trivial. In particular, at the energy region of the charmed meson, such pole terms may become significant and cannot be neglected. It has been shown in Ref.~\cite{Cao:2023csx} that the internal conversion processes can play a crucial role in some of those $D^0\to VV$ channels. This pushes us to include these contributions and make efforts on estimating their magnitude with some accessible constraints.

{\it Type-I pole terms}
    
    Since for the type-I pole amplitude, the photon emission occurs before the weak transition. Hence, the intermediate state can be either a vector meson for the parity-conserving case or an axial vector meson for the parity-violating one. This intermediate state will propagate virtually until it weakly decays into a vector meson $V$. The type-I pole amplitude is then given by
\begin{align}
    \begin{split}
        \mathcal{M}_{I}^{\text{(P)}}(D\to V\gamma)=\sum_n\langle V|H_{W,2\to2}^{\text{(P)}}\frac{i}{m_V^2-m_{D_n^*}^2+im_{D_n^*}\Gamma_{D_n^*}}|D_n^*\rangle\langle D_n^*|H_{EM}|D\rangle,
    \end{split}
\end{align}
where $H_{W,2\to2}^{(\text{P})}$ is the weak effective Hamiltonian for the $2\to2$ internal conversion process.

\begin{itemize}
    \item PC transition amplitude
    \begin{align}
        \begin{split}
            i\mathcal{M}_{I}^{\text{(PC)}}&=(i\mathcal{M}_{D\to V^{\prime}D^*})G_{V^{\prime}}(i\mathcal{M}_{V^{\prime}\gamma})G_{D^*}(i\mathcal{M}_{D^*V})\\
            &=(ig_{DD^*V^{\prime}}\epsilon_{\alpha\beta\mu\lambda}ip_{V^{\prime}}^{\alpha}ip_{D^*}^{\beta})\frac{-i(g^{\mu\nu}-\frac{p_{V^{\prime}}^{\mu}p_{V^{\prime}}^{\nu}}{p_{V^{\prime}}^2})}{p_{V^{\prime}}^2-m_{V^{\prime}}^2+im_{V^{\prime}}\Gamma_{V^{\prime}}}(\frac{iem_{V^{\prime}}^2}{f_{V^{\prime}}}\varepsilon_{\gamma}^{\nu})\frac{-i(g^{\lambda\delta}-\frac{p_{D^{*}}^{\lambda}p_{D^{*}}^{\delta}}{p_{D^{*}}^2})}{p_{D^{*}}^2-m_{D^{*}}^2+im_{D^{*}}\Gamma_{D^{*}}}(ig_{D^*V}^{(\text{PC})}\varepsilon_{V}^{\delta})\\
            &\times\delta(p_{V^{\prime}}-p_{\gamma})\delta(p_{D^*}-p_{V})\\
            &=i\epsilon_{\alpha\beta\mu\lambda}p_{\gamma}^{\alpha}p_V^{\beta}\varepsilon_{\gamma}^{\mu}\varepsilon_V^{\lambda}(g_{DD^*V^{\prime}}\frac{em_{V^{\prime}}^2}{f_{V^{\prime}}}\frac{-i}{p_{\gamma}^2-m_{V^{\prime}}^2+im_{V^{\prime}}\Gamma_{V^{\prime}}}g_{D^*V}^{\text{(PC)}}\frac{-i}{m_V^2-m_{D^{*}}^2+im_{D^{*}}\Gamma_{D^{*}}})\\
            &=ig_{DV\gamma}^{\text{(PC)}}\epsilon_{\alpha\beta\mu\lambda}p_{\gamma}^{\alpha}p_V^{\beta}\varepsilon_{\gamma}^{\mu}\varepsilon_V^{\lambda},
        \end{split}
    \end{align}
    where $g_{DV\gamma}^{\text{(PC)}}=g_{DD^*(1^-)V^{\prime}}\frac{em_{V^{\prime}}^2}{f_{V^{\prime}}}G_{V^{\prime}}g_{D^*V}^{\text{(PC)}}\mathbb{G}_{V,D^*}$, $G_V\equiv\frac{-i}{p_{\gamma}^2-m_{V}^2+im_{V}\Gamma_{V}}=\frac{-i}{-m_{V}^2+im_{V}\Gamma_{V}}$, and $\mathbb{G}_{V,V^{\prime}}\equiv\frac{-i}{m_V^2-m_{V^{\prime}}^2+im_{V^{\prime}}\Gamma_{V^{\prime}}}$.
    \item PV transition amplitude
    \begin{align}
        \begin{split}
            i\mathcal{M}_I^{(\text{(PV)})}&=(i\mathcal{M}_{D\to V^{\prime}D^*})G_{V^{\prime}}(i\mathcal{M}_{V^{\prime}\gamma})G_{D^*}(i\mathcal{M}_{D^*V})\\
            &=(ig_{DD^*V^{\prime}})\frac{-i(g^{\mu\nu}-\frac{p_{V^{\prime}}^{\mu}p_{V^{\prime}}^{\nu}}{p_{V^{\prime}}^2})}{p_{V^{\prime}}^2-m_{V^{\prime}}^2+im_{V^{\prime}}\Gamma_{V^{\prime}}}(\frac{iem_{V^{\prime}}^2}{f_{V^{\prime}}}\varepsilon_{\gamma}^{\nu})\frac{-i(g^{\mu\delta}-\frac{p_{D^{*}}^{\mu}p_{D^{*}}^{\delta}}{p_{D^{*}}^2})}{p_{D^{*}}^2-m_{D^{*}}^2+im_{D^{*}}\Gamma_{D^{*}}}(ig_{D^*V}^{(\text{PV})}\varepsilon_V^{\delta})\\
            &\times \delta(p_{V^{\prime}}-p_{\gamma})\delta(p_{D^*}-p_V)\\
            &=(ig_{DD^*V^{\prime}})\frac{-i}{p_{\gamma}^2-m_{V^{\prime}}^2+im_{V^{\prime}}\Gamma_{V^{\prime}}}(\frac{iem_{V^{\prime}}^2}{f_{V^{\prime}}}\varepsilon_{\gamma}^{\mu})\frac{-i}{m_{V}^2-m_{D^{*}}^2+im_{D^{*}}\Gamma_{D^{*}}}(ig_{D^*V}^{(\text{PV})}\varepsilon_{V\mu})\\
            &=-ig_{DV\gamma}^{(\text{PV})}\varepsilon_{\gamma}^{\mu}\varepsilon_{V\mu},
        \end{split}
    \end{align}
    where $p_{\gamma}\cdot\varepsilon_{\gamma}=p_V\cdot\varepsilon_V=0$, $g_{DV\gamma}^{(\text{PV})}=g_{DD^*(1^+)V^{\prime}}\frac{em_{V^{\prime}}^2}{f_{V^{\prime}}}G_{V^{\prime}}g_{D^*V}^{(\text{PV})}\mathbb{G}_{V,D^*}$, and functions $G_{V^{\prime}}$ and $\mathbb{G}_{V,D^*}$ are the same as the above.
\end{itemize}

To estimate the coupling strength $g_{DD^*(1^-)V}$, the following Lagrangians are invoked,
\begin{align}
    \begin{split}
        \mathcal{L}_{VD\bar{D}^*}&=g_{VD\bar{D}^*}\epsilon_{\alpha\beta\mu\nu}\partial^{\alpha}V^{\beta}\partial^{\mu}{\bar{D}^{*\nu}}D+h.c.,\\
        \mathcal{L}_{\psi D\bar{D}^*}&=g_{\psi D\bar{D}^*}\epsilon_{\alpha\beta\mu\nu}\partial^{\alpha}\psi^{\beta}\partial^{\mu}{\bar{D}^{*\nu}}D+h.c.,
    \end{split}
\end{align}
where $V$ is the light vector meson field. In the chiral and heavy quark limits, the following relations can be obtained \cite{Cheng:2004ru,Colangelo:2003sa}
\begin{align}
    \begin{split}
        g_{VD\bar{D}^*}&=\sqrt{2}\lambda g_V,\ \ \ \ \ g_V=\frac{m_{\rho}}{f_\pi},\\
        g_{J/\psi D\bar{D}^{*}}&=\frac{g_{J/\psi D\bar{D}}}{\tilde{M}_D},\ \ \ \ \tilde{M}_D=\sqrt{m_Dm_{D^*}},
    \end{split}
\end{align} 
where $\lambda$ is commonly taken as $\lambda=0.56$ GeV$^{-1}$ and $f_{\pi}=132$ MeV is the pion decay constant. The coupling $g_{J/\psi D\bar{D}}$ can be extracted by the VMD model and its value is $g_{J/\psi D\bar{D}}=7.44$~\cite{Oh:2007ej}.

Unfortunately, we know little about the property of the $D^0$ couplings to the axial-vector meson $D_1$ and a neutral vector meson. Therefore, we do not consider the parity-violating contributions of the type-I amplitudes here.  But we will put an estimate of the uncertainties from the PV pole terms.

{\it  Type-II pole terms}
    
For the Type-II pole terms, the intermediate state is either a scalar for the PV transition or a pseudoscalar meson for the PC one. In the VMD scenario the intermediate state will propagate virtually until it decays to a pair of vector mesons. Then, the flavor-neutral one will transit into a photon. The amplitude can be written in a similar way as the type-I amplitude: 
\begin{align}
    \begin{split}
        \mathcal{M}_{II}^{\text{(P)}}(D\to V\gamma)=\sum_n\langle V|H_{EM}\frac{i}{m_D^2-m_{P_n}^2+im_{P_n}\Gamma_{P_n}}|P_n\rangle\langle P_n|H_{W,2\to2}^{\text{(P)}}|D\rangle.
    \end{split}
\end{align}

\begin{itemize}
    \item PC transition amplitude
    \begin{align}
        \begin{split}
            i\mathcal{M}_{II}^{(\text{PC})}&=(i\mathcal{M}_{DP})G_P(i\mathcal{M}_{P\to VV^{\prime}})G_{V^{\prime}}(iM_{V^{\prime}\gamma})\\
            &=(ig_{DP}^{(\text{PC})})\frac{i}{p_{P}^2-m_P^2+im_P\Gamma_P}(ig_{PVV^{\prime}}\epsilon_{\alpha\beta\mu\delta}ip_{V^{\prime}}^{\alpha}ip_V^{\beta}\varepsilon_V^{\delta})\frac{-i(g^{\mu\nu}-\frac{p_{V^{\prime}}^{\mu}p_{V^{\prime}}^{\nu}}{p_{V^{\prime}}^2})}{p_{V^{\prime}}^2-m_{V^{\prime}}^2+im_{V^{\prime}}\Gamma_{V^{\prime}}}(\frac{iem_{V^{\prime}}^2}{f_{V^{\prime}}}\varepsilon_{\gamma}^{\nu})\\
            &\times\delta(p_{V^{\prime}}-p_{\gamma})\delta(p_D-p_P)\\
            &=i\epsilon_{\alpha\beta\mu\delta}p_{\gamma}^{\alpha}p_V^{\delta}\varepsilon_{\gamma}^{\mu}\varepsilon_V^{\delta}(g_{DP}^{(\text{PC})}\frac{i}{m_{D}^2-m_P^2+im_P\Gamma_P}g_{PVV^{\prime}}\frac{-i}{p_{\gamma}^2-m_{V^{\prime}}^2+im_{V^{\prime}}\Gamma_{V^{\prime}}}\frac{em_{V^{\prime}}^2}{f_{V^{\prime}}})\\
            &=ig_{DV\gamma}^{(\text{PC})}\epsilon_{\alpha\beta\mu\delta}p_{\gamma}^{\alpha}p_V^{\delta}\varepsilon_{\gamma}^{\mu}\varepsilon_V^{\delta},
        \end{split}
    \end{align}
    where $g_{DV\gamma}^{(\text{PC})}=g_{DP}^{(\text{PC})}\mathbb{G}_{D,P}g_{PVV^{\prime}}G_{V^{\prime}}\frac{em_{V^{\prime}}^2}{f_{V^{\prime}}}$.
    \item PV transition amplitude
    \begin{align}
        \begin{split}
            i\mathcal{M}_{II}^{(\text{PV})}(D\to V\gamma)&=(i\mathcal{M}_{DS})G_S(i\mathcal{M}_{S\to VV^{\prime}})G_{V^{\prime}}(iM_{V^{\prime}\gamma})\\
            &=(ig_{DS}^{(\text{PV})})\frac{i}{p_{S}^2-m_S^2+im_S\Gamma_S}(ig_{SVV^{\prime}}\varepsilon_V^{\mu})\frac{-i(g^{\mu\nu}-\frac{p_{V^{\prime}}^{\mu}p_{V^{\prime}}^{\nu}}{p_{V^{\prime}}^2})}{p_{V^{\prime}}^2-m_{V^{\prime}}^2+im_{V^{\prime}}\Gamma_{V^{\prime}}}(\frac{iem_{V^{\prime}}^2}{f_{V^{\prime}}}\varepsilon_{\gamma}^{\nu})\\
            &\times\delta(p_{V^{\prime}}-p_{\gamma})\delta(p_D-p_S)\\
            &=-i\varepsilon_{\gamma}^{\mu}\varepsilon_{V\mu}(g_{DS}^{(\text{PV})}\frac{i}{m_{D}^2-m_S^2+im_S\Gamma_S}g_{SVV^{\prime}}\frac{-i}{p_{\gamma}^2-m_{V^{\prime}}^2+im_{V^{\prime}}\Gamma_{V^{\prime}}}\frac{em_{V^{\prime}}^2}{f_{V^{\prime}}})\\
            &=-ig_{DV\gamma}^{(\text{PV})}\varepsilon_{\gamma}^{\mu}\varepsilon_{V\mu},
        \end{split}
    \end{align}
    where $g_{DV\gamma}=g_{DS}^{(\text{PV})}\mathbb{G}_{D,S}g_{SVV^{\prime}}G_{V^{\prime}}\frac{em_{V^{\prime}}^2}{f_{V^{\prime}}}$.
\end{itemize}

For the PV transitions, we again encounter the problem of lacking the coupling information for the intermediate scalar meson transits into the vector meson pair. While it is impossible to provide a quantified prescription here, we will estimate the magnitude of uncertainties arising from this process.

We collect the PC amplitudes from the IC processes which includes both type-I and type-II:
\begin{eqnarray}
\mathcal{M}_{\text{IC}}^{(\text{PC})}&=&\mathcal{M}_{I(b)}^{(\text{PC})}+\mathcal{M}_{II(c)}^{(\text{PC})} \ ,
\end{eqnarray}
with
\begin{eqnarray}
        \mathcal{M}_{I(a)}^{(\text{PC})}&\equiv &\sum_{D_n^*(1^-)}\langle V(1^-)|H_{W,2\to 2}^{\text{(PC)}}\frac{i}{m_{V}^2-m_{D_n^{*0}}^2+im_{D_n^{*0}}\Gamma_{D_n^{*0}}}|D_n^{*0}(1^-)\rangle\langle D_n^{*0}(1^-)|H_{EM}|D^0(0^-)\rangle,\\
        \mathcal{M}_{II(b)}^{(\text{PC})}&\equiv&\sum_{P_n(0^-)}\langle V(1^-)|H_{EM}\frac{i}{m_{D}^2-m_{P_n}^2+im_{P_n}\Gamma_{P_n}}|P_n(0^-)\rangle\langle P_n(0^-)|H_{W,2\to2}^{\text{(PC)}}|D^0(0^-) \rangle,
\end{eqnarray}
where a complete set of intermediate meson states $D_n^*$ ($P_n$) with quantum numbers $1^{-}$ ($0^{-}$) have been included in process of type-I and type-II, respectively. $\langle P(V)|H_{W,2\to 2}^{\text{(PC)}}|D^{(*)}\rangle$ is the weak transition matrix element which can be parametrized out as listed in Tab. \ref{table: 2 to 2 weak matrix elements}. 

To obtain the $EM$ transition matrix element, we describe the electromagnetic vertices in the VMD model. Taking a pair of charm mesons $D^0\bar{D}^{*0}$ couplings to the photon $\gamma$ as an example, the vertex can be written as
\begin{align}
    \begin{split}
        g_{D^0D^{*0}\gamma}={i}g_{\rho^0 D^0\bar{D}^{*0}}\frac{em_{\rho^0}^2}{f_{\rho^0}}G_{\rho^0}+{i}g_{\omega D^0\bar{D}^{*0}}\frac{em_{\omega}^2}{f_{\omega}}G_{\omega}+ig_{J/\psi D^0\bar{D}^{*0}}\frac{em_{J/\psi}^2}{f_{J/\psi}}{R}G_{J/\psi},
    \end{split}
\end{align}
where $R$ is an SU(4) flavor symmetry breaking parameter and it distinguishes the production of a $c\bar{c}$ from that of $u\bar{u}$ ($d\bar{d}$) and takes a value of $0.3$.  We present detailed expressions for all electromagnetic vertices in the VMD model in Appendix~\ref{appendix: expressions of EM couplings} and their values are listed in Tab.~\ref{tab: VMD couplings}. 

\begin{table}
    \centering
    \caption{Weak transition matrix elements of the internal conversion processes. The coefficient includes flavor factor and spin sign. $\alpha_P$ is the mixing angle of the $\eta$ and $\eta^{\prime}$ on the quark-flavor basis and it takes a value of $42^{\circ}$.  Coupling $g_{W(\text{IC}^)}^{(\text{P})}$ can be extracted by calculating the $2\to 2$ internal conversion process in the NRCQM.}
    \label{table: 2 to 2 weak matrix elements}
    \begin{tabular}{cc|cc}
        \hline\hline
        Procrsses                        &$\langle H_{W,2\to2}^{(\text{PC})}\rangle$                                           &Procrsses                        &$\langle H_{W,2\to2}^{(\text{PC})}\rangle$\\
        \hline
        $D^0\to\bar{K}^{0}$             &$-V_{cs}V_{ud}g_{W\text{(IC)}}^{\text{(PC)}}$                                        &$D^*(2007)^0\to\bar{K}^{*0}$     &$V_{cs}V_{ud}g_{W\text{(IC)}}^{\text{(PC)}}$\\
        \hline
        $D^0\to\eta(s\bar{s})$           &$\text{sin}\alpha_PV_{cs}V_{us}g_{W\text{(IC)}}^{\text{(PC)}}$                       &$D^*(2007)^0\to\phi$             &$V_{cs}V_{us}g_{W\text{(IC)}}^{\text{(PC)}}$\\
        \hline
        $D^0\to\eta^{\prime}(s\bar{s})$  &$-\text{cos}\alpha_PV_{cs}V_{us}g_{W\text{(IC)}}^{\text{(PC)}}$                      &$D^*(2007)^0\to\rho^0$           &$-\frac{1}{\sqrt{2}}V_{cd}V_{ud}g_{W\text{(IC)}}^{\text{(PC)}}$\\
        \hline
        $D^0\to\eta(d\bar{d})$           &$-\frac{1}{\sqrt{2}}\text{cos}\alpha_PV_{cd}V_{ud}g_{W\text{(IC)}}^{\text{(PC)}}$    &$D^*(2007)^0\to\omega$           &$\frac{1}{\sqrt{2}}V_{cd}V_{ud}g_{W\text{(IC)}}^{\text{(PC)}}$\\
        \hline
        $D^0\to\eta^{\prime}(d\bar{d})$  &$-\frac{1}{\sqrt{2}}\text{sin}\alpha_PV_{cd}V_{ud}g_{W\text{(IC)}}^{\text{(PC)}}$    &-&-\\
        \hline
        $D^0\to\pi^0(d\bar{d})$          &$\frac{1}{\sqrt{2}}V_{cd}V_{ud}g_{W\text{(IC)}}^{\text{(PC)}}$                       &-&-\\
        \hline\hline
    \end{tabular}
\end{table}

\begin{table}
    \centering
    \caption{The effective couplings for the electromagnetic vertices. The values extracted from the VMD model are listed in the second column and their corresponding modules are presented in the round brackets. In the last column the signs of the values are determined in the quark model. The couplings $g_{K^*K^*\gamma}(g_{\rho\rho\gamma})$ and $g_{KK\gamma}(g_{\pi\pi\gamma})$ are treated as pure QED couplings. Thus their coupling strengths will be given by the charge of the hadron. Note that $e\simeq 0.33$.}
    \label{tab: VMD couplings}
    \begin{tabular}{c|c|c}
        \hline\hline
        Electromagnetic couplings                    &Values in VMD (Magnitude)    &Experimental values\\
        \hline
        $g_{K^+K^{*+}\gamma}$ [$\text{GeV}^{-1}$]    &$-0.288-0.063i\ (0.294)$         &$-0.253\pm 0.012$\\
        $g_{K^0K^{*0}\gamma}$ [$\text{GeV}^{-1}$]    &$0.369+0.062i\ (0.374)$         &$0.385\pm 0.016$\\
        \hline
        $g_{K^{*+}K^{*+}\gamma}$                     &$0.324+0.036i\ (0.326)$        &$e$\\
        $g_{K^{*0}K^{*0}\gamma}$                     &$-0.046-0.034i\ (0.057)$        &$0$\\
        \hline
        $g_{K^{+}K^{+}\gamma}$                       &$0.324+0.036i\ (0.326)$        &$e$\\
        $g_{K^{0}K^{0}\gamma}$                       &$-0.046-0.034i\ (0.057)$        &$0$\\
        \hline
        $g_{\pi^+\rho^{+}\gamma}$ [$\text{GeV}^{-1}$]    &$-0.205-0.002i\ (0.205)$     &$-0.219\pm 0.012$\\
        $g_{\pi^0\rho^{0}\gamma}$ [$\text{GeV}^{-1}$]    &$-0.205-0.002i\ (0.205)$     &$-0.222\pm 0.019$\\
        \hline
        $g_{\rho^{+}\rho^{+}\gamma}$                     &$0.347+0.066i\ (0.354)$        &$e$\\
        $g_{\rho^{0}\rho^{0}\gamma}$                     &$0$                            &$0$\\
        \hline
        $g_{\pi^{+}\pi^{+}\gamma}$                       &$0.347+0.066i\ (0.354)$        &$e$\\
        $g_{\pi^{0}\pi^{0}\gamma}$                       &$0$                          &$0$\\            
        \hline
        $g_{\eta\phi\gamma}$ [$\text{GeV}^{-1}$]         &$-0.192-0.0008i\ (0.192)$    &$-0.209\pm 0.002$\\
        \hline
        $g_{\eta^{\prime}\phi\gamma}$ [$\text{GeV}^{-1}$]&$0.213+0.0009i\ (0.213)$     &$0.217\pm 0.004$\\
        \hline
        $g_{\eta\rho^0\gamma}$ [$\text{GeV}^{-1}$]       &$-0.488-0.093i\ (0.496)$     &$-0.478\pm 0.017$\\
        \hline
        $g_{\eta^{\prime}\rho^0\gamma}$ [$\text{GeV}^{-1}$] &$-0.440-0.084i\ (0.447)$   &$-0.434\pm 0.003$\\
        \hline
        $g_{\eta\omega\gamma}$ [$\text{GeV}^{-1}$]       &$-0.152-0.002i\ (0.152)$      &$-0.136\pm 0.006$\\
        \hline
        $g_{\eta^{\prime}\omega\gamma}$ [$\text{GeV}^{-1}$] &$-0.137-0.002i\ (0.137)$   &$-0.134\pm 0.002$\\
        \hline
        $g_{\pi^0\omega\gamma}$ [$\text{GeV}^{-1}$]       &$-0.657-0.125i\ (0.669)$     &$-0.707\pm 0.011$\\
        \hline
        $g_{D^0D^{*0}\gamma}$ [$\text{GeV}^{-1}$]         &$-0.389-0.053i\ (0.393)$   &$>-3.297$\\
        \hline\hline
    \end{tabular}
\end{table}

The intermediate states of the pole terms contribute differently in these four radiative decay processes. Since the inclusion of all the intermediate states are impractical we only consider the ground state as the leading contribution. Namely, for the PC case, only $D^*(2007)^0$ will contribute to the type-I pole term of all four decays; $\bar{K}^0$ will contribute to the type-II pole term in $D^0\to \bar{K}^{*0}\gamma$; both $\eta$ and $\eta^{\prime}$ will contribute to the type-II pole terms in $D^0\to\phi\gamma$, $D^0\to\rho^0\gamma$, and $D^0\to\omega\gamma$, and in addition, $\pi^0$ will contribute to type-II pole terms in $D^0\to\rho^0\gamma$ and $D^0\to\omega\gamma$.

\begin{table}
    \centering
    \caption{The tree and pole amplitudes for different processes and the unit is $10^{-8}$ GeV$^{-1}$. Amplitudes $\mathcal{A}_{I(b)}^{\text{PC}}$, $\mathcal{A}_{II(c1)}^{\text{PC}}$, $\mathcal{A}_{II(c2)}^{\text{PC}}$, $\mathcal{A}_{II(c3)}^{\text{PC}}$ and $\mathcal{A}_{II(c4)}^{\text{PC}}$ are given by the parity-conserving intermediate states $D^*(2007)^0$, $\bar{K}^{0}$, $\eta$, $\eta^{\prime}$ and $\pi^0$, respectively.}
    \label{tab: tree and pole amplitudes}
    \begin{tabular}{cccccccccc}
        \hline\hline
        Mode                      &$\mathcal{A}_{T\text{(a)}}^{\text{(PV)}}$       &$\mathcal{A}_{T\text{(a)}}^{\text{(PC)}}$          &$\mathcal{A}_{I\text{(b)}}^{\text{(PC)}}$        &$\mathcal{A}_{II\text{(c1)}}^{\text{(PC)}}$        &$\mathcal{A}_{II\text{(c2)}}^{\text{(PC)}}$      &$\mathcal{A}_{II\text{(c3)}}^{\text{(PC)}}$    &$\mathcal{A}_{II\text{(c4)}}^{\text{(PC)}}$   &$\mathcal{A}_I^{(\text{PC})}+\mathcal{A}_{II}^{(\text{PC})}$\\
        \hline
        $\bar{K}^{*0}\gamma$     &$1.20-8.17i$                                     &$-1.13+7.68i$                                      &$0.79-5.76i$                                     &$-0.93+5.57i$                                       &0                                                &0                                              &0                                             &$-0.14+0.19i$\\
        \hline
        $\phi\gamma$             &$0.34-2.30i$                                     &$-0.29+1.97i$                                      &$0.19-1.43i$                                     &0                                                  &$-0.002+0.45$                                    &$-0.003+0.69i$                                  &0                                             &$0.19-0.28i$\\
        \hline
        $\rho^0\gamma$           &$-0.16+1.39i$                                    &$0.16-1.40i$                                       &$-0.12+0.87i$                                    &0                                                  &$0.16-0.89i$                                     &$0.17-0.89i$                                    &$-0.005+0.47i$                                &$0.21-0.45i$\\
        \hline
        $\omega\gamma$           &$0.0004+0.32i$                                   &$-0.0005-0.32i$                                    &$0.12-0.87i$                                     &0                                                  &$0.003-0.27i$                                    &$0.003-0.28i$                                  &$-0.28+1.46i$                                  &$-0.15+0.03i$\\
        \hline\hline        
    \end{tabular}
\end{table}

\begin{table}
    \centering
    \caption{The b.r.s of the tree contributions shown in Fig. \ref{fig: D-V gamma}(a) and pole term contributions shown in Fig. \ref{fig: D-V gamma}(b) and (c) in units of $10^{-5}$ for the decay $D^0\to V\gamma$ ($V=\bar{K}^{*0},\ \phi,\ \rho^0,\ \omega$).}
    \label{table: tree and pole b.r.}
    \begin{tabular}{c|c|c|c|c|c}
        \hline\hline
        \multicolumn{2}{c|}{b.r.}                                           &$\bar{K}^{*0}\gamma$   &$\phi\gamma$     &$\rho^0\gamma$   &$\omega\gamma$\\
        \hline
        \multicolumn{2}{c|}{Experimental data\ \cite{ParticleDataGroup:2022pth}}  &$41\pm 7$              &$2.81\pm 0.19$   &$1.82\pm 0.32$   &$<24$\\
        \hline
        \multirow{3}{*}{Tree}           &PC                             &$11.03\pm5$                      &$0.55\pm0.23$    &$0.45\pm0.16$    &$0.023\pm 0.020$\\
                                        &PV                             &$12.53\pm4$                      &$0.75\pm0.25$    &$0.45\pm0.13$    &$0.023\pm 0.016$\\
                                        &PC+PV                          &$23.57\pm6$                      &$1.30\pm0.34$    &$0.90\pm0.21$    &$0.047\pm 0.026$\\
        \hline
        Pole terms                      &PC                             &$0.011$                          &$0.016$          &$0.055$          &$0.005$\\
        \hline
        \multicolumn{2}{c|}{Tree(PC+PV)\ +\ Pole terms(PC)}             &$22.80$                          &$1.14$           &$1.23$           &$0.046$\\
        \hline\hline
    \end{tabular}
\end{table}

We summarize some of the main features of the pole term contributions as follows:
\begin{itemize}
    \item  Since only the ground states are considered, the IC pole terms will mainly contribute to the real part of the transition amplitude due to the narrowness of these intermediate states. Moreover, the type-I and type-II amplitudes have opposite signs and will cancel each other. For the same intermediate $D^0\to VV'$ processes, the cancellation turns to be apparent if an infinity mass limit is taken for the heavy quark, e.g. $m_{D_n^{*0}}>>m_V$ and $m_{D^0}>>m_{P_n}$.

    \item The relative sign between the weak transition matrix element $\langle V(1^-)|H_{W,2\to2}^{(\text{PC})}|D_n^{*0}\rangle$ for the type-I pole terms and $\langle P_n(0^-)|H_{W,2\to2}^{(\text{PC})}|D^0(0^-)\rangle$ is explicitly fixed by the SU(3) flavor symmetry as shown in Tab.~\ref{table: 2 to 2 weak matrix elements}. 
    
    \item In Tab.~\ref{tab: VMD couplings} it shows that the charmed meson coupling to the photon $g_{D^0D^{*0}\gamma}$ in the type-I pole terms, is at the same order of magnitude as that of the light meson couplings to the photon in the type-II pole terms. 

    \item The above properties lead to the cancellation between the amplitudes of the two types of pole terms even in  the physical $D$ meson mass region. We list the amplitudes of the pole terms in Tab. \ref{tab: tree and pole amplitudes} and compare them with the corresponding tree amplitudes. One can see that the exclusive type-I or type-II amplitudes can be sizeable. However, by comparing the exclusive contributions from the tree and the sum of the pole terms, we see that the cancellation has led to rather small effects on the b.r.s as shown in Tab.~\ref{table: tree and pole b.r.}. We argue that the parity-violating intermediate states in the IC processes have the similar behaviors. As a  result of the cancellation, the IC pole terms turn to be much smaller than the CS process (Fig. \ref{fig: QuarkLevel diagrams}(a)).

\end{itemize}

\subsection{Loop amplitudes in the VMD model}

In the VMD model the mass of the intermediate vector meson pairs are close to that of $D^0$. Nevertheless, the mass of the $K^*\bar{K}^*$ is almost degenerate with that of $\phi\rho^0$ and $\phi\omega$. It suggests that FSIs via the $VV\to VV$ rescatterings can be significant. In addition, other predominant intermediate processes which can rescatter into the flavor-neutral $VV$ channel can also contribute if the quantum numbers are allowed. This is recently investigated by Ref.~\cite{Cao:2023csx}. To some extent, the FSIs are anticipated in $D^0\to V\gamma$ in the VMD model.

The typical FSI processes can be illustrated by Fig. \ref{fig: D-V gamma} (d), where the DE process as the intermediate channel should be crucial. To benefit from the study of Ref.~\cite{Cao:2023csx} where the $D^0\to VV$ couplings have been extracted, we define the leading short-distance couplings $g_{W\text{(SD)}}^{\text{(P)}}\equiv g_{W(\text{DE})}^{(\text{P})}+e^{i\theta}g_{W(\text{IC})}^{(\text{P})}$ where $\theta=\pi$ is taken as determined in Ref.~\cite{Cao:2023csx}. Then, we obtain the weak couplings $g_{W(\text{SD})}^{(\text{PC})}=2.0\times 10^{-6}\ \text{GeV}^{-1}$ and $g_{W(\text{SD})}^{(\text{PV})}=2.4\times 10^{-6}\ \text{GeV}$ for the CF decay $D^0\to K^{*-}\rho^+$, and $g_{W(\text{SD})}^{(\text{PC})}=1.5\times 10^{-6}\ \text{GeV}^{-1}$ and $g_{W(\text{SD})}^{(\text{PV})}=4.5\times 10^{-6}\ \text{GeV}$ for the SCS decay $D^0\to K^{*+}K^{*-}$. These vertex couplings will be adopted for the calculations of the FSIs via  Fig.~\ref{fig: D-V gamma}(d).

The triangle loop amplitudes illustrated by Fig. \ref{fig: D-V gamma}(d) can also be calculated in the VMD model. The loop amplitudes can reduce to an effective coupling which contributes to the $DV\gamma$ coupling in the end. Within the triangle loops the vertices for the photon couplings to the kaon (pion) and/or $K^*\ (\rho)$ pairs can be described by the VMD model. Taking the $K^{*+}K^-$ coupling to the photon $\gamma$ as an example, the photon can couple to the intermediate $\rho^0$, $\omega$, and $\phi$ mesons, via the following amplitude, 
\begin{align}
    \begin{split}
        g_{K^{*+}K^+\gamma}=\sum_{q=u,s}\sum_{V^{\prime}=\rho^0, \omega, \phi}\langle (q\bar{q})_{1^{--}}|V^{\prime}\rangle ig_{V^{\prime}VP}\frac{em_{V^{\prime}}^2}{f_{V^{\prime}}}G_{V^{\prime}},
    \end{split}
\end{align}
where $V$ and $P$ stand for the initial $K^{*+}$ and pseudoscalar meson $K^-$, while $V^{\prime}$ stands for the intermediate vector mesons, $\rho^0$, $\omega$, and $\phi$, to which the photon can couple with a strength of the decay constant $e/f_{V^{\prime}}$; $\langle (q\bar{q})_{1^{--}}|V^{\prime}\rangle$ is a favor factor given by the decomposition of the $q\bar{q}$ into flavor eigenstate of the intermediate vector mesons. For instance, 
\begin{align}
    \begin{split}
        \langle (u\bar{u})_{1^{--}}|V\rangle &=\langle\frac12(u\bar{u}-d\bar{d})+\frac12(u\bar{u}+d\bar{d}) |V \rangle={\frac{1}{\sqrt{2}}}\langle (\rho^0+\omega)|V\rangle.
    \end{split}
\end{align}

We present the detailed expressions by the VMD model for the electromagnetic vertices in Appendix~\ref{appendix: expressions of EM couplings}. In those equations the ground-state vector meson decay constants $e/f_V\ (V=\phi,\ \rho^0,\ \omega)$ are extracted with the data for $V\to e^+e^-$~\cite{ParticleDataGroup:2022pth}.

Since the hadronic vertices can be related by the SU(3) flavor symmetry their relative strengths and phases can be fixed. There are five types of hadronic coupling vertices in the loop amplitudes, i.e.,
$VPP$, $VVP$, $VVV$, $SPP$ and $SVV$ for which the corresponding effective Lagrangians are as follows:
\begin{align}\label{eff-lagrangian}
    \mathcal{L}_{VPP}&=ig_{VPP}Tr[(P\partial_{\mu}P-\partial_{\mu}PP)V^{\mu}],\\
    \mathcal{L}_{VVP}&=g_{VVP}\epsilon_{\alpha\beta\mu\nu}Tr[\partial^{\alpha}V^{\mu}\partial^{\beta}V^{\nu}P],\\
    \mathcal{L}_{VVV}&=ig_{VVV}Tr[(\partial_{\mu}V_{\nu}-\partial_{\nu}V_{\mu})V^{\mu}V^{\nu}],\\
    \mathcal{L}_{PPS}&=g_{PPS}Tr[PPS],\\
    \mathcal{L}_{VVS}&=g_{VVS}Tr[VVS].
  \end{align} 
   where the vector ($V$), pseudoscalar ($P$), and scalar ($S$) fields as the SU(3) flavor multiplets are listed as follows, respectively, 
   \begin{equation}\label{su3-vector}
     V=\left(
       \begin{array}{ccc}
         \frac{\omega+\rho^{0}}{\sqrt{2}} & \rho^{+} & {K^{*}}^+ \\
         \rho^{-} &  \frac{\omega-\rho^{0}}{\sqrt{2}} & K^{*0} \\
         K^{*-}& \bar{K}^{*0} & \phi \\
      \end{array}
     \right),
   \end{equation}
   \begin{equation}\label{su3-pseudo}
     P=\left(
       \begin{array}{ccc}
         \frac{\sin\alpha_P \eta'+ \cos \alpha_P\eta+\pi^0}{\sqrt{2}} & \pi^{+}& K^{+}\\
         \pi^{-} & \frac{ \sin\alpha_P \eta'+ \cos \alpha_P\eta-\pi^0}{\sqrt{2}}& K^{0} \\
         K^{-} & \bar{K^{0}}& \cos\alpha_P \eta'-\sin\alpha_P \eta \\
       \end{array}
   \right),
   \end{equation}
   and
   \begin{equation}\label{su3-scalar}
     S=\begin{pmatrix}
       \frac{ \sigma + a_0(980)}{\sqrt{2}} &a_0^+                               & \kappa^+ \\
         a_0^-                           &  \frac{ \sigma-a_0(980)}{\sqrt{2}} & \kappa^0 \\
         \kappa^-                        &   \bar{\kappa}^0                   &f_0(980)
     \end{pmatrix},
     \end{equation}
where the ideal mixing are adopted betwee $\omega(=(u\bar{u}+d\bar{d})/\sqrt{2})$ and $\phi(=s\bar{s})$.
 
With the effective Lagrangians in Eq.~(\ref{eff-lagrangian}), we can write down the loop transition amplitude for Fig. \ref{fig: D-V gamma}(d). We use the notation $\tilde{\mathcal{I}}[(\text{P}), M_1, M_3, (M_2)]$ to denote the loop amplitude. Namely, the intermediate $M_1$ and $M_3$ rescatter into $V\gamma$ by exchanging $M_2$, and (P) ($=$(PC) or
(PV)) indicates the PC or PV property of the corresponding amplitude. The masses and $4$-vector momenta of these internal particles are denoted by $(m_1,\ m_3,\ m_2)$ and $(p_1,\ p_3,\ p_2)$, respectively. The $4$-vector momenta of the initial-state meson $D^0$, final-state photon, and vector meson are labeled as $p_D$, $p_{\gamma}$, and $p_V$,
respectively. The polarizations of the final-state photon and vector meson are $\varepsilon_{\gamma}$ and $\varepsilon_V$, respectively. For instance, the intermediate $K^{*+}K^{*-}$ rescattering amplitudes through the  triangle loops by exchanging $\mathbb{K}$ ($K$ or $K^*$) can be expressed as follows:
\begin{align}
    \begin{split}
        i\mathcal{M}^{(\text{P})}_{loop}=\sum_{\mathbb{K}}\tilde{\mathcal{I}}[(\text{P}),\ K^{*+},\ K^{*-},\ (\mathbb{K})],
    \end{split}
\end{align}
where the sum is over the contributing meson loops indicated by the different exchanged mesons. 

Taking the PC loop transition $[(\text{PC});\ K^{*},\ \bar{K}^{*},\ (K)]$ as an example, the loop integral is:
\begin{eqnarray}
       &&\tilde{\mathcal{I}}[(\text{PC});\ K^{*},\ \bar{K}^{*},\ (K)]\nonumber\\
       & =&\int\frac{d^4p_1}{(2\pi)^4}V_{1\mu\nu}D^{\mu\mu^{\prime}}(K^{*})V_{2\mu^{\prime}}D(K)V_{3\nu^{\prime}}D^{\nu\nu^{\prime}}(\bar{K}^{*})\mathcal{F}(p_i^2),\label{eq: im1}
    \end{eqnarray}
where the vertex functions have compact forms as follows:
\begin{align}
    \begin{split}
        V_{1\mu\nu}&=-ig_{W(\text{SD})}^{(\text{P})} V_{cs}V_{us}\epsilon_{\alpha\beta\mu\nu}p_1^{\alpha}p_3^{\beta},\\
        V_{2\mu^{\prime}}&=ig_{K^*\bar{K}\gamma}\epsilon_{\alpha_1\beta_1\mu^{\prime}\delta}p_1^{\alpha_1}p_{\gamma}^{\beta_1}\varepsilon_{\gamma}^{*\delta},\\
        V_{3\nu^{\prime}}&=ig_{V\bar{K}^*K}\epsilon_{\alpha_2\beta_2\nu^{\prime}\lambda}p_3^{\alpha_2}p_{V}^{\beta_2}\varepsilon_{V}^{*\lambda} \ ,
        \label{eq: vertex-1P}
    \end{split}
\end{align}
where functions $D^{\mu\mu^{\prime}}(K^*)$ and $D(K)$ are the propagators for $K^{*}$ and $K$, respectively, with 4-vector momentum $p$, i.e.,
\begin{align}
    \begin{split}
        D^{\mu\mu^{\prime}}(K^*)&=\frac{-i(g^{\mu\mu^{\prime}}-\frac{p^{\mu}p^{\mu^{\prime}}}{p^2})}{p^2-m_{K^*}^2+i\epsilon},\\
        D(K)&=\frac{i}{p^2-m_{K}^2+i\epsilon}.
    \end{split}
\end{align}
We note that all the strong vertex couplings involving the light pseudoscalar ($P$) and vector ($V$) meson couplings, i.e., $g_{VPP}$, $g_{VVP}$ and $g_{VVV}$, have been extracted by Refs. \cite{Cheng:2021nal, Cheng:2023lov}, such as $g_{V\bar{K}^*K}$ in Eq. (\ref{eq: vertex-1P}). In addition, the electromagnetic couplings are extracted by the VMD model and listed in Tab. \ref{tab: VMD couplings}, such as $g_{K^*\bar{K}\gamma}$ in Eq. (\ref{eq: vertex-1P}).

In order to cut off the ultra-violet (UV) divergence in the loop integrals, we include a commonly-adopted form factor to regularize the integrand:
\begin{align}
    \mathcal{F}(p_i^2)=\prod_i(\frac{\Lambda_i^2-m_i^2}{\Lambda_i^2-p_i^2}),
\end{align}
where $\Lambda_i\equiv m_i+\alpha\Lambda_{\text{QCD}}$ with $\Lambda_{\text{QCD}}=220 \ \text{MeV}$ and $\alpha=1\sim 2$ as the cut-off parameter.

\section{Numerical results and discussions}

The final results for the $D^0$ weak radiative decays are given by the interfering contributions from both the short-distance and long-distance dynamics. As studied in Sec.~\ref{subsect-tree} the short-distance contributions include the tree-level transitions which are calculated in the NRCQM. The main uncertainties arises from the PV intermediate pole terms for which the couplings cannot be well constrained. However, considering that the type-I and type-II terms cancel each other in this case, we expect that the neglect of the PV pole terms would not produce significant uncertainties. For the long-distance contributions, the FSIs via the triangle loops can be accommodated by the VMD framework consistently, and the main uncertainties arises from the UV cut-off parameter for the triangle loop integrals. However, since the FSIs in $D^0\to VV$ have been quantitatively studied recently in Ref.~\cite{Cao:2023csx}, a consistent range of the cut-off parameter has been determined. We adopt it as a prediction for this analysis. The non-trivial aspect is that the SU(3) flavor symmetry has determined all the relative phases among the transition amplitudes, which means that we are obliged to describe the existing data within the commonly adopted values for the cut-off parameter.

In Tab.~\ref{tab: tree and pole amplitudes} the amplitudes of the short-distance processes have been calculated, while in Tab.~\ref{table: tree and pole b.r.} the corresponding b.r.s are also obtained. One sees a systematic underestimate of the experimental data~\cite{ParticleDataGroup:2022pth}, and it calls for additional mechanism to account for the deficits. 

The contributions from the long-distance processes via the FSIs is shown by Tab.~\ref{tab: each loop diagram b.r.}, where the contributions from exclusive triangle loop transitions to the b.r.s are listed. It shows that the vector meson exchange terms are dominant in all the triangle loop transitions. This is a feature that we found in $D^0\to VV$~\cite{Cao:2023csx}. It is also interesting to note that the FSI corrections are mainly to the PV channel, while the PC corrections are negligibly small in comparison with the tree-level transitions. It is understandable that the PV coupling for $D^0$ to the intermediate $VV$ is an $S$ wave while the PC coupling is a $P$ wave which will bring significant suppressions near threshold. In addition, the yields of the loop integrals in the PV channel are also found larger than in the PC channel due to the different structures of the integrands.

As shown in Tab.~\ref{tab: each loop diagram b.r.} the exclusive b.r.s from the FSIs indicate some sensitivities to the cut-off parameter for $\alpha=1$ and $2$. This appears to be the main source of the theoretical uncertainties. Combining the tree and triangle loop amplitudes together, we obtain the full results for the b.r.s for these four decay channels,  i.e. $D^0\to V\gamma$ ($V=\bar{K}^{*0}$, $\phi$, $\rho^0$, $\omega$), with $\delta=0$ as the natural phase in Tab.~\ref{tab: tree and loop b.r.}. We also list the experimental data from the BaBar~\cite{BaBar:2008kjd}, Belle Collaboration~\cite{Belle:2016mtj}, and the PDG average~\cite{ParticleDataGroup:2022pth} as a comparison. It shows that the measurements of $D^0\to\bar{K}^{*0}\gamma$ by these two experiments turn out to be quite different, while their measurements of $D^0\to\phi\gamma$ are consistent with each other. Meanwhile, the channel of $D^0\to\rho^0\gamma$ was only measured by Belle~\cite{Belle:2016mtj} and an upper limit for $D^0\to\omega\gamma$ was set by the CLEO Collaboration~\cite{CLEO:1998mtp}. By best describing $D^0\to\phi\gamma$ and $\rho^0\gamma$ with $\alpha=1.3\pm 0.13$, we find that the calculated b.r. for $D^0\to\bar{K}^{*0}\gamma$ is consistent with the averaged value of the BaBar and Belle measurements, and the b.r. for $D^0\to\omega\gamma$ is much smaller than the experimental upper limit. In comparison with $D^0\to\rho^0\gamma$ the suppressed b.r. of $D^0\to\omega\gamma$ can be well understood by the destructive interference between the two tree-level amplitudes in Tab.~\ref{Table: tree-level amplitude} as discussed earlier. 

In Tab.~\ref{tab: each loop diagram b.r.} we also include other theoretical calculations~\cite{Burdman:1995te,Biswas:2017eyn,Fajfer:1998dv,Shen:2013oua,deBoer:2017que,Asthana:1990kr,Bajc:1994ui,Dias:2017nwd} in the literature as a comparison. One can see that most of these existing results have significantly underestimated the data except for Ref.~\cite{Shen:2013oua} which has given the results with the correct order of magnitude.

\begin{figure}
    \centering
    \subfigure[$D^0\to \bar{K}^{*0}\gamma$]{\includegraphics[width=2.5in]{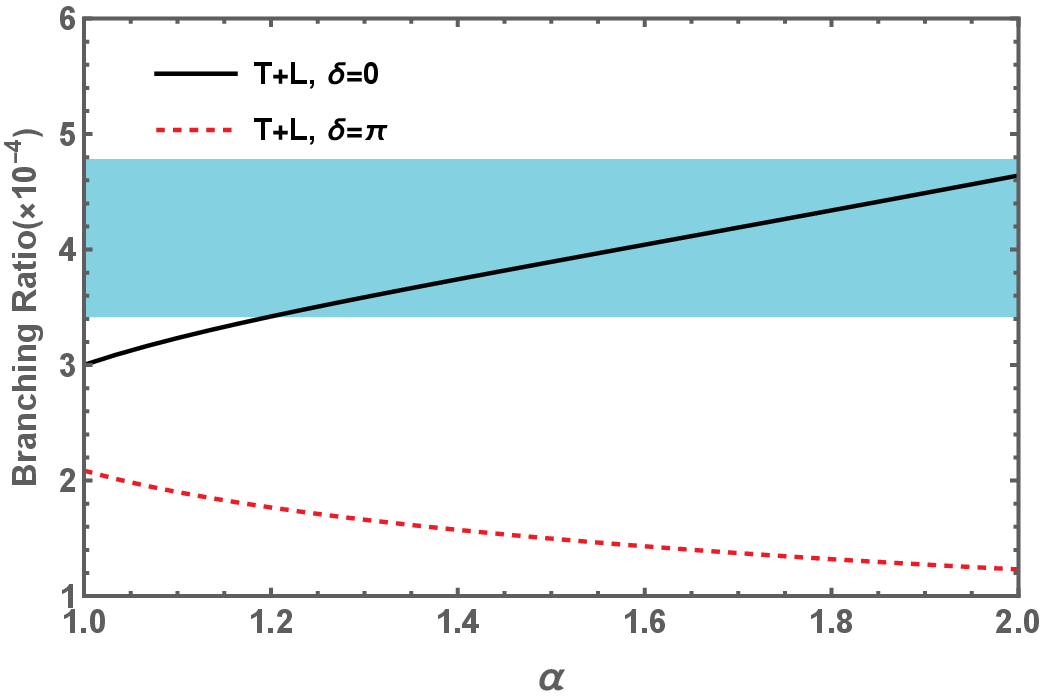}}
    \hspace{0.9mm}
    \subfigure[$D^0\to\phi\gamma$]{\includegraphics[width=2.5in]{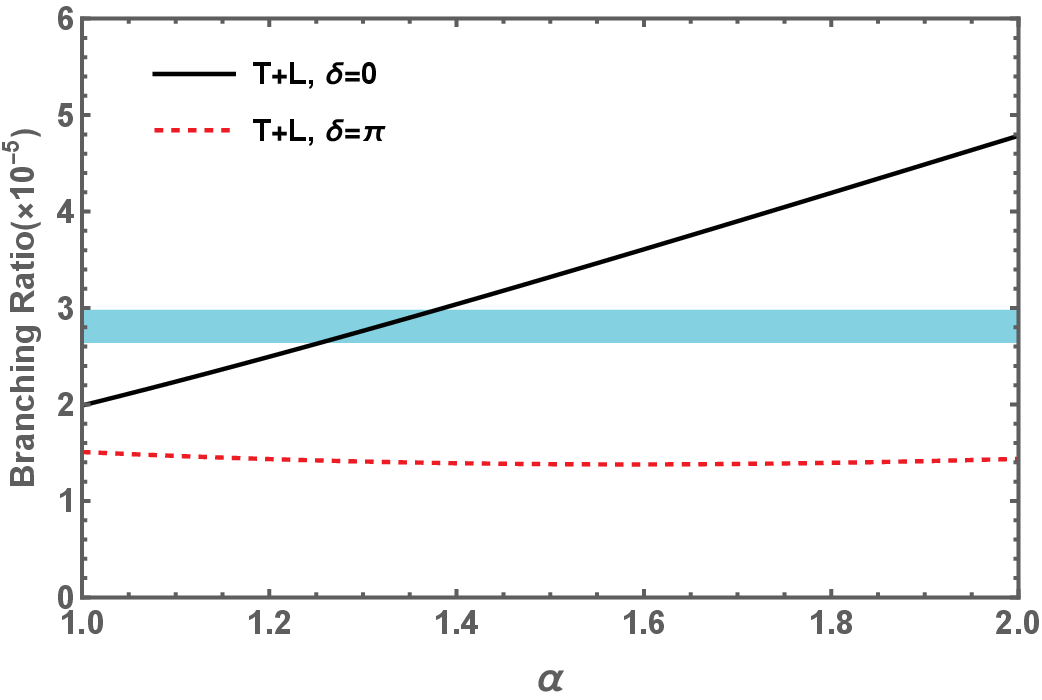}}
    \hspace{0.9mm}
    \subfigure[$D^0\to\rho^0\gamma$]{\includegraphics[width=2.5in]{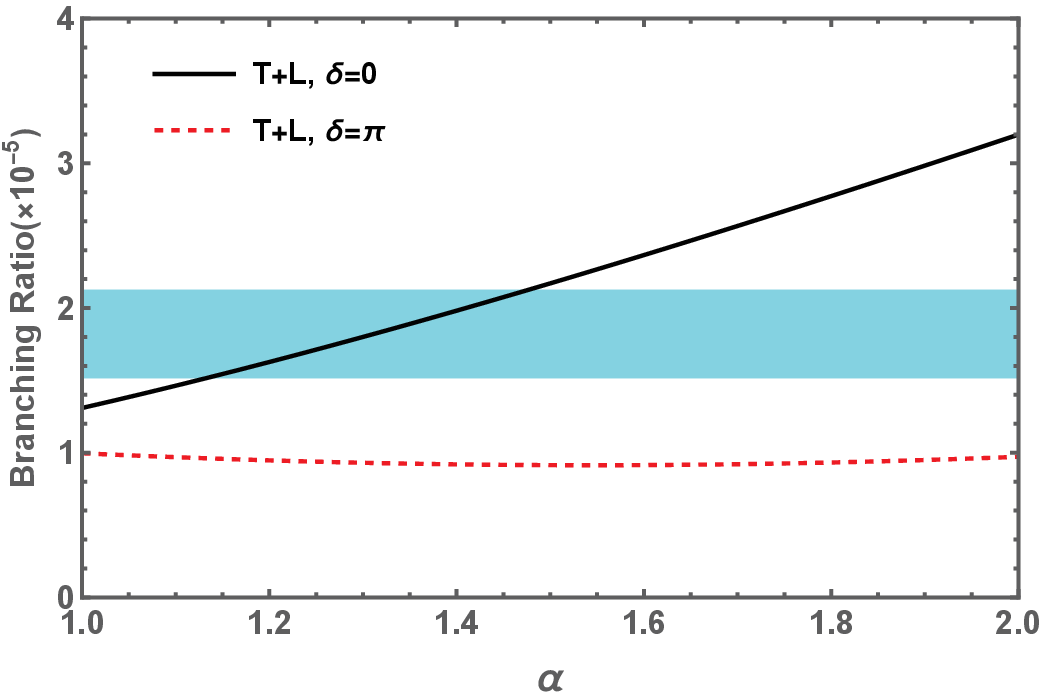}}
    \hspace{0.9mm}
    \subfigure[$D^0\to\omega\gamma$]{\includegraphics[width=2.55in]{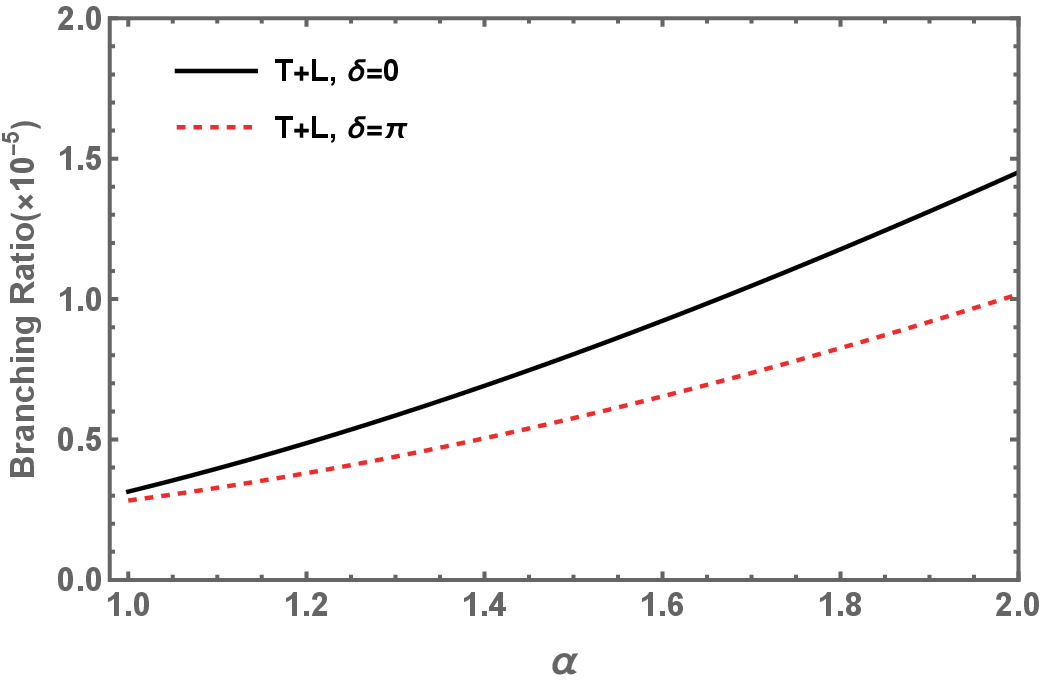}}
    \caption{The dependence of the b.r.s on the cut-off parameter $\alpha$ for the four radiative weak decays of charm meson $D^0$. ``T+L" denotes the full calculations including the tree (Fig. \ref{fig: D-V gamma}(a)) and loop (Fig. \ref{fig: D-V gamma}(d)) contributions. $\delta$ is the relative phase of the tree diagrams and loop diagrams. The horizontal bands are the PDG averages of the corresponding experimental data~\cite{ParticleDataGroup:2022pth}.}
    \label{fig: lineshape b.r. with alpha}
\end{figure}

To see more clearly the role played by the FSIs, we plot the cut-off parameter $\alpha$ dependence of the b.r.s in Fig.~\ref{fig: lineshape b.r. with alpha}, where ``T+L" denotes the full calculations including both the tree (Fig. \ref{fig: D-V gamma}(a)) and triangle loop (Fig. \ref{fig: D-V gamma}(d)) contributions. We also show two interfering patterns between the tree and triangle loop amplitudes denoted by $\delta=0$ and $\pi$, respectively. Namely, $\delta=0$ means a natural phase determined by the SU(3) flavor symmetry while $\delta=\pi$ denotes an opposite extreme.  The horizontal bands are the PDG average~\cite{ParticleDataGroup:2022pth} of the experimental measurements. For these four decay channels, i.e. $D^0\to V\gamma$ ($V=\bar{K}^{*0}$, $\phi$, $\rho^0$, $\omega$), we find that the data for $D^0\to \bar{K}^{*0}$, $\phi$, $\rho^0$ can be accounted for within a range of $\alpha=1.3\pm 0.13$ with $\delta=0$. In contrast, the results with $\delta=\pi$ cannot describe these three channels simultaneously. This confirms our anticipation of the natural sign due to the SU(3) flavor symmetry in this analysis.

\begin{table}
    \centering 
    \caption{Calculated b.r.s of each type of the hadronic loop diagrams in units of $10^{-5}$ with the cut-off parameter $\alpha=1$ and $\alpha=2$.}
    \label{tab: each loop diagram b.r.}
    \begin{tabular}{c|c|ccc|ccc}
        \hline\hline
        \multirow{2}{*}{Diagrams}                             &\multirow{2}{*}{Decay channels}    &\multicolumn{3}{c}{$\alpha=1$}                                              &\multicolumn{3}{c}{$\alpha=2$}\\
        \cline{3-8}
                                                              &                                   &(PC)                     &(PV)                   &(PC)+(PV)                 &(PC)                    &(PV)                    &(PC)+(PV)\\
        \hline
        \multirow{3}{*}{$[K^{*+},\ K^{*-},\ (K^+)]$}          &$\phi\gamma$                       &$5.79\times 10^{-5}$     &$1.14\times 10^{-2}$   &$1.15\times 10^{-2}$      &$5.33\times 10^{-4}$    &$5.24\times 10^{-2}$    &$5.29\times 10^{-2}$\\
                                                              &$\rho^0\gamma$                     &$1.72\times 10^{-5}$     &$5.21\times 10^{-3}$   &$5.23\times 10^{-3}$      &$1.97\times 10^{-4}$    &$2.79\times 10^{-2}$    &$2.81\times 10^{-2}$\\
                                                              &$\omega\gamma$                     &$1.74\times 10^{-5}$     &$5.22\times 10^{-3}$   &$5.23\times 10^{-3}$      &$1.98\times 10^{-4}$    &$2.79\times 10^{-2}$    &$2.81\times 10^{-2}$\\
        \hline
        \multirow{3}{*}{$[K^{*+},\ K^{*-},\ (K^{*+})]$}       &$\phi\gamma$                       &$3.29\times 10^{-3}$     &$0.61$                 &$0.61$                    &$2.23\times 10^{-2}$    &$2.15$                  &$2.17$\\
                                                              &$\rho^0\gamma$                     &$1.10\times 10^{-3}$     &$0.18$                 &$0.18$                    &$9.93\times 10^{-3}$    &$0.84$                  &$0.85$\\
                                                              &$\omega\gamma$                     &$1.11\times 10^{-3}$     &$0.18$                 &$0.18$                    &$9.92\times 10^{-3}$    &$0.85$                  &$0.86$\\
        \hline
        $[K^{*-},\ \rho^+,\ (K^-)]$                            &$\bar{K}^{*0}\gamma$              &$1.39\times 10^{-3}$     &$2.73\times 10^{-2}$   &$2.87\times 10^{-2}$      &$1.11\times 10^{-2}$    &$8.95\times 10^{-2}$    &$0.10$\\
        \hline
        $[K^{*-},\ \rho^+,\ (K^{*-})]$                         &$\bar{K}^{*0}\gamma$              &$7.24\times 10^{-2}$     &$1.18$                 &$1.25$                    &$0.44$                  &$3.60$                  &$4.04$\\
        \hline
        $[\rho^+,\ K^{*-},\ (\pi^+)]$                          &$\bar{K}^{*0}\gamma$              &$1.18\times 10^{-4}$     &$1.42\times 10^{-2}$   &$1.43\times 10^{-2}$      &$2.33\times 10^{-3}$    &$5.98\times 10^{-2}$    &$6.21\times 10^{-2}$\\
        \hline
        $[\rho^+,\ K^{*-},\ (\rho^+)]$                         &$\bar{K}^{*0}\gamma$              &$0.12$                   &$1.78$                 &$1.91$                    &$0.72$                  &$5.20$                  &$5.92$\\
        \hline\hline
    \end{tabular}
\end{table}

\begin{table}
    \centering
    \caption{Calculated b.r.s containing both tree and loop contributions in units of $10^{-5}$ for the four radiative weak decays $D^0\to V\gamma$ ($V=\bar{K}^{*0}$, $\phi$, $\rho^0$, $\omega$). The uncertainties are given by $\alpha=1.3\pm 0.13$.}
    \label{tab: tree and loop b.r.}
    \begin{tabular}{c|cccc}
        \hline\hline
                                                            &$D^0\to\bar{K}^{*0}\gamma$                   &$D^0\to\phi\gamma$                              &$D^0\to\rho^0\gamma$          &$D^0\to\omega\gamma$\\
        \hline
        \multirow{2}{*}{Experimental data}                  &$32.8\pm2.0\pm2.7$ \cite{BaBar:2008kjd}      &$2.78\pm 0.32\pm 0.27$ \cite{BaBar:2008kjd}    
        &\multirow{2}{*}{$1.77\pm 0.30\pm 0.07$ \cite{Belle:2016mtj}}&\multirow{2}{*}{$<24$ \cite{CLEO:1998mtp}}\\
                                                            &$46.6\pm 2.1\pm 2.1$ \cite{Belle:2016mtj}    &$2.76\pm 0.19\pm 0.10$ \cite{Belle:2016mtj}\\
        \hline
 PDG average~\cite{ParticleDataGroup:2022pth}  & $41\pm 7$       &$2.81\pm 0.19$   &$1.82\pm 0.32$   &$<24$ \\
        \hline
        Burdman \cite{Burdman:1995te}                          &$7\sim 12$ &$0.1\sim 3.4$ &$0.1\sim 0.5$ &$\sim 0.2$\\
        \hline
        Biswas \cite{Biswas:2017eyn}                                        &$4.6\sim 18$ &$0.48\sim 0.64$ &$0.512\sim 1.8$ &$0.32\sim 0.9$\\
        \hline
        Fajfer \cite{Fajfer:1998dv}                           &$6\sim 36$ &$0.4\sim 1.9$ &$0.1\sim 1$ &$0.1\sim 0.9$\\
        \hline
        Shen \cite{Shen:2013oua}                             &$19_{-6-1}^{+7+1}$ &$3.2_{-1.0-0.0}^{+1.3+0.3}$ &$1.1_{-0.4-0.1}^{+0.4+0.1}$ &$0.75_{-0.25-0.04}^{+0.30+0.05}$\\
        \hline
        de Boer \cite{deBoer:2017que}                           &$2.6\sim 46$ &$0.24\sim 2.8$ &$0.041\sim 1.17$ &$0.042\sim 1.12$\\
        \hline
        Asthana \cite{Asthana:1990kr}                          &0.86 &- &- &-\\
        \hline
        Bajc \cite{Bajc:1994ui}                             &$28\sim 65$ &- &- &-\\
        \hline
        Dias \cite{Dias:2017nwd}                             &$15.5\sim 34.4$ &- &- &-\\
        \hline
        This work                      &$35.9^{+2.0}_{-2.2}$   &$2.76_{-0.35}^{+0.36}$   &$1.79_{-0.22}^{+0.24}$  &$0.58_{-0.13}^{+0.14}$\\
        \hline\hline        
    \end{tabular}
\end{table}

\section{Summary}

In this work we have systematically studied the Cabibbo-favored and singly Cabibbo-suppressed $D^0$ weak radiative decays into $V\gamma$ in the framework of the VMD model. By distinguishing the short and long-distance transition mechanisms we demonstrate that the long-distance FSIs have played a crucial role in the understanding of the $D^0\to V\gamma$ decays. In particular, the intermediate $K^{*+}K^{*-}$ rescatterings by exchanging a vector meson into $V\gamma$ account for most of the long-distance contributions. Since the same mechanism also plays a crucial role in the understanding of the hadronic weak decays of $D^0\to VV$, this study can be regarded as a self-consistent examination of the long-distance dynamics in the $D^0$ weak decays. We also emphasize that the extraction of the tree-level amplitudes is a consistent way to take into account the SU(3) flavor symmetry breaking effects, which arises from the constituent quark mass differences. This mechanism may be investigated in other decay channels given that more and more data for the $D$ mesons have been accumulated in experiment, and it can help us gain better insights into the near-threshold dynamics via the weak transitions.

\section*{Acknowledgement}
This work is supported, in part, by the National Natural Science Foundation of China (Grant No. 12235018), DFG and NSFC funds to the Sino-German CRC 110 ``Symmetries and the Emergence of Structure in QCD" (NSFC Grant No. 12070131001, DFG Project-ID 196253076), National Key Basic Research Program of China under Contract No. 2020YFA0406300, and Strategic Priority Research Program of Chinese Academy of Sciences (Grant No. XDB34030302).

\appendix

\begin{appendix}

\section{Weak effective Hamiltonian for internal conversion processes }
\label{appendix:}

The non-relativistic form of the weak effective Hamiltonian for the $2\to 2$ ($q_1\bar{q}_2\to q_1^{\prime}\bar{q}_2^{\prime}$) internal conversion process can be explicitly reduced as
\begin{align}
    \begin{split}
        H_{W,2\to 2}^{(\text{PC})}&=\frac{G_F}{\sqrt{2}}V_{q_1q_1^{\prime}}V_{q_2q_2^{\prime}}\frac{1}{(2\pi)^3}\sum_{i\neq j}\hat{\alpha}_i^{(-)}\hat{\beta}_j^{(+)}\delta^2(\boldsymbol{p}_i^{\prime}+\boldsymbol{p}_j^{\prime}-\boldsymbol{p}_i-\boldsymbol{p}_j)\\
        &\times(\langle\bar{s_i}^{\prime}|I|\bar{s_i}\rangle\langle{s_j^{\prime}}|I|{s_j}\rangle-\langle\bar{s_i}^{\prime}|\boldsymbol{\sigma}_i|\bar{s_i}\rangle\cdot\langle{s_j^{\prime}}|\boldsymbol{\sigma}_j|{s_j}\rangle),\\
        H_{W,2\to 2}^{(\text{PV})}&=\frac{G_F}{\sqrt{2}}V_{q_1q_1^{\prime}}V_{q_2q_2^{\prime}}\frac{1}{(2\pi)^3}\sum_{i\neq j}\hat{\alpha}_i^{(-)}\hat{\beta}_j^{(+)}\delta^2(\boldsymbol{p}_i^{\prime}+\boldsymbol{p}_j^{\prime}-\boldsymbol{p}_i-\boldsymbol{p}_j)\\
        &\times \{ -(\langle\bar{s_i}^{\prime}|\boldsymbol{\sigma}_i|\bar{s_i}\rangle\langle{s_j^{\prime}}|I|{s_j}\rangle-\langle\bar{s_i}^{\prime}|I|\bar{s_i}\rangle\cdot\langle{s_j^{\prime}}|\boldsymbol{\sigma}_j|{s_j}\rangle)\cdot\left[\left(\frac{\boldsymbol{p}_i}{2m_i}-\frac{\boldsymbol{p}_j}{2m_j}\right)+\left(\frac{\boldsymbol{p}_i^{\prime}}{2m_i^{\prime}}-\frac{\boldsymbol{p}_j^{\prime}}{2m_j^{\prime}}\right)\right]\\
        &\times i(\langle\bar{s_i}^{\prime}|\boldsymbol{\sigma}_i|\bar{s_i}\rangle\times\langle{s_j^{\prime}}|\boldsymbol{\sigma}_j|{s_j}\rangle)\cdot\left[\left(\frac{\boldsymbol{p}_i^{\prime}}{2m_i^{\prime}}+\frac{\boldsymbol{p}_j^{\prime}}{2m_j^{\prime}}\right)-\left(\frac{\boldsymbol{p}_i}{2m_i}+\frac{\boldsymbol{p}_j}{2m_j}\right)\right]
        \},
    \end{split}
\end{align}
where $s_i$ and $m_i$ are the spin and mass of the $i$-th quark ($\bar{s}_i$ stands for the spin of particle $i$ which is an antiquark), respectively; the subscripts $i$ and $j$ indicate the quarks experiencing the weak interaction; $\hat{\alpha}_i^{(-)}$ and $\hat{\beta}_j^{(+)}$ are the flavor-changing operators, e.g., $\hat{\alpha}_i^{(-)}c_j=\delta_{ij}s_j$, $\hat{\beta}_j^{(+)}u_i=\delta_{ij}d_j$ for Cabibbo-favored process.

\section{The detailed  expressions of the  electromagnetic coupling in the VMD model}
\label{appendix: expressions of EM couplings}

In this Appendix, we present the detailed expressions for the electromagnetic couplings extracted in the VMD model. In the following equations, $R$ takes a value of 0.8 to distinguish the production of an $s\bar{s}$ from $u\bar{u}$ and $d\bar{d}$ due to the SU(3) flavor symmetry breaking:
\begin{align}
    \begin{split}
    g_{K^+K^{*+}\gamma}&=i(g_{\rho^0K^{*+}K^-}\frac{em_{\rho^0}^2}{f_{\rho^0}}G_{\rho^0}+g_{\omega K^{*+}K^-}\frac{em_{\omega}^2}{f_{\omega}}G_{\omega}+g_{\phi K^{*+}K^-}\frac{em_{\phi}^2}{f_{\phi}}RG_{\phi})\\
    &=\frac{i}{\sqrt{2}}(g_{VVP}\frac{em_{\rho^0}^2}{f_{\rho^0}}G_{\rho^0}+g_{VVP}\frac{em_{\omega}^2}{f_{\omega}}G_{\omega})+ig_{VVP}\frac{em_{\phi}^2}{f_{\phi}}RG_{\phi},\\
    g_{K^0K^{*0}\gamma}&=i(g_{\rho^0K^{*0}\bar{K}^{*0}}\frac{em_{\rho^0}^2}{f_{\rho^0}}G_{\rho^0}+g_{\omega K^{*0}\bar{K}^{*0}}\frac{em_{\omega}^2}{f_{\omega}}G_{\omega}+g_{\phi K^{*0}\bar{K}^{*0}}\frac{em_{\phi}^2}{f_{\phi}}RG_{\phi})\\
    &=\frac{i}{\sqrt{2}}(-g_{VVP}\frac{em_{\rho^0}^2}{f_{\rho^0}}G_{\rho^0}+g_{VVP}\frac{em_{\omega}^2}{f_{\omega}}G_{\omega})+ig_{VVP}\frac{em_{\phi}^2}{f_{\phi}}RG_{\phi},\\
    g_{K^{*+}K^{*+}\gamma}&=i(g_{\rho^0K^{*+}K^{*-}}\frac{em_{\rho^0}^2}{f_{\rho^0}}G_{\rho^0}+g_{\omega K^{*+}K^{*-}}\frac{em_{\omega}^2}{f_{\omega}}G_{\omega}+g_{\phi K^{*+}K^{*-}}\frac{em_{\phi}^2}{f_{\phi}}RG_{\phi})\\
    &=\frac{i}{\sqrt{2}}(-g_{VVV}\frac{em_{\rho^0}^2}{f_{\rho^0}}G_{\rho^0}-g_{VVV}\frac{em_{\omega}^2}{f_{\omega}}G_{\omega})+ig_{VVV}\frac{em_{\phi}^2}{f_{\phi}}RG_{\phi},\\
    g_{K^{*0}{K}^{*0}\gamma}&=i(g_{\rho^0K^{*0}\bar{K}^{*0}}\frac{em_{\rho^0}^2}{f_{\rho^0}}G_{\rho^0}+g_{\omega K^{*0}\bar{K}^{*0}}\frac{em_{\omega}^2}{f_{\omega}}G_{\omega}+g_{\phi K^{*0}\bar{K}^{*0}}\frac{em_{\phi}^2}{f_{\phi}}RG_{\phi})\\
    &=\frac{i}{\sqrt{2}}(g_{VVV}\frac{em_{\rho^0}^2}{f_{\rho^0}}G_{\rho^0}-g_{VVV}\frac{em_{\omega}^2}{f_{\omega}}G_{\omega})+ig_{VVV}\frac{em_{\phi}^2}{f_{\phi}}RG_{\phi},\\
    g_{K^{+}K^{+}\gamma}&=i(g_{\rho^0K^{+}K^{-}}\frac{em_{\rho^0}^2}{f_{\rho^0}}G_{\rho^0}+g_{\omega K^{+}K^{-}}\frac{em_{\omega}^2}{f_{\omega}}G_{\omega}+g_{\phi K^{+}K^{-}}\frac{em_{\phi}^2}{f_{\phi}}RG_{\phi})\\
    &=\frac{i}{\sqrt{2}}(-g_{VPP}\frac{em_{\rho^0}^2}{f_{\rho^0}}G_{\rho^0}-g_{VPP}\frac{em_{\omega}^2}{f_{\omega}}G_{\omega})+ig_{VPP}\frac{em_{\phi}^2}{f_{\phi}}RG_{\phi},\\
    g_{K^{0}K^{0}\gamma}&=i(g_{\rho^0K^{0}\bar{K}^{0}}\frac{em_{\rho^0}^2}{f_{\rho^0}}G_{\rho^0}+g_{\omega K^{0}\bar{K}^{0}}\frac{em_{\omega}^2}{f_{\omega}}G_{\omega}+g_{\phi K^{0}\bar{K}^{0}}\frac{em_{\phi}^2}{f_{\phi}}RG_{\phi})\\
    &=\frac{i}{\sqrt{2}}(g_{VPP}\frac{em_{\rho^0}^2}{f_{\rho^0}}G_{\rho^0}-g_{VPP}\frac{em_{\omega}^2}{f_{\omega}}G_{\omega})+ig_{VPP}\frac{em_{\phi}^2}{f_{\phi}}RG_{\phi}.\\
    \end{split}
\end{align}
\begin{align}
    \begin{split}
    g_{\pi^+\rho^+\gamma}&=ig_{\omega\rho^+\pi^-}\frac{em_{\omega}^2}{f_{\omega}}G_{\omega}=i\sqrt{2}g_{VVP}\frac{em_{\omega}^2}{f_{\omega}}G_{\omega},\\
    g_{\pi^0\rho^0\gamma}&=ig_{\omega\rho^0\pi^0}\frac{em_{\omega}^2}{f_{\omega}}G_{\omega}=i\sqrt{2}g_{VVP}\frac{em_{\omega}^2}{f_{\omega}}G_{\omega},\\
    g_{\pi^0\omega\gamma}&=ig_{\pi^0\omega\rho^0}\frac{em_{\rho^0}^2}{f_{\rho^0}}G_{\rho^0}=i\sqrt{2}g_{VVP}\frac{em_{\rho^0}^2}{f_{\rho^0}}G_{\rho^0},\\
    g_{\rho^+\rho^+\gamma}&=ig_{\rho^0\rho^+\rho^-}\frac{em_{\rho^0}^2}{f_{\rho^0}}G_{\rho^0}=i(-\sqrt{2}g_{VVV})\frac{em_{\rho^0}^2}{f_{\rho^0}}G_{\rho^0},\\
    g_{\rho^0\rho^0\gamma}&=0,\\
    g_{\pi^+\pi^+\gamma}&=ig_{\rho^0\pi^+\pi^-}\frac{em_{\rho^0}^2}{f_{\rho^0}}G_{\rho^0}=i(-\sqrt{2}g_{VPP})\frac{em_{\rho^0}^2}{f_{\rho^0}}G_{\rho^0},\\
    g_{\pi^0\pi^0\gamma}&=0.\\
    \end{split}
\end{align}

\begin{align}
    \begin{split}
        g_{\eta\phi\gamma}&=ig_{\eta\phi\phi}\frac{em_{\phi}^2}{f_{\phi}}{R}G_{\phi}=i(-2\text{sin}\alpha_Pg_{VVP})\frac{em_{\phi}^2}{f_{\phi}}{R}G_{\phi},\\
        g_{\eta^{\prime}\phi\gamma}&=ig_{\eta^{\prime}\phi\phi}\frac{em_{\phi}^2}{f_{\phi}}{R}G_{\phi}=i(2\text{cos}\alpha_Pg_{VVP})\frac{em_{\phi}^2}{f_{\phi}}{R}G_{\phi},\\
        g_{\eta\rho^0\gamma}&=ig_{\eta\rho^0\rho^0}\frac{em_{\rho^0}^2}{f_{\rho^0}}G_{\rho^0}=i(\sqrt{2}\text{cos}\alpha_Pg_{VVP})\frac{em_{\rho^0}^2}{f_{\rho^0}}G_{\rho^0},\\
        g_{\eta^{\prime}\rho^0\gamma}&=ig_{\eta^{\prime}\rho^0\rho^0}\frac{em_{\rho^0}^2}{f_{\rho^0}}G_{\rho^0}=i(\sqrt{2}\text{sin}\alpha_Pg_{VVP})\frac{em_{\rho^0}^2}{f_{\rho^0}}G_{\rho^0},\\
        g_{\eta\omega\gamma}&=ig_{\eta\omega\omega}\frac{em_{\omega}^2}{f_{\omega}}G_{\omega}=i(\sqrt{2}\text{cos}\alpha_Pg_{VVP})\frac{em_{\omega}^2}{f_{\omega}}G_{\omega},\\
        g_{\eta^{\prime}\omega\gamma}&=ig_{\eta^{\prime}\omega\omega}\frac{em_{\omega}^2}{f_{\omega}}G_{\omega}=i(\sqrt{2}\text{sin}\alpha_Pg_{VVP})\frac{em_{\omega}^2}{f_{\omega}}G_{\omega}.
    \end{split}
\end{align}

\section{The flavor SU(3) relationship of the coupling constants}
\label{appendix: SU(3) relationship}

The relative strengths and phases of the vector-pseudoscalar couplings can be
fixed and expressed by overall coupling coefficients, i.e. $g_{VPP}$, $g_{VVP}$ and $g_{VVV}$, considering the SU(3) flavor symmetry:
\begin{itemize}
    \item $VPP$ vertices
    \begin{align}
        \begin{split}
            g_{\phi K^+K^-}&=-g_{\phi K^-K^+}=g_{\phi K^0 \bar{K}^0}=-g_{\phi \bar{K}^0K^0}=g_{VPP},\\
            g_{\omega K^+K^-}&=-g_{\omega K^-K^+}=g_{\omega K^0 \bar{K}^0}=-g_{\omega \bar{K}^0K^0}=-\frac{1}{\sqrt{2}}g_{VPP},\\
            g_{\rho^0 K^+K^-}&=-g_{\rho^0 K^-K^+}=-g_{\rho^0 K^0 \bar{K}^0}=g_{\rho^0 \bar{K}^0K^0}=-\frac{1}{\sqrt{2}}g_{VPP}.
        \end{split}
    \end{align}
    \item $VVP$ vertices
    \begin{align}
        \begin{split}
            g_{\phi K^{*+}K^-}&=g_{\phi K^{*-}K^+}=g_{\phi K^{*0}\bar{K}^0}=g_{\phi \bar{K}^{*0}K^0}=g_{VVP},\\
            g_{\omega K^{*+}K^-}&=g_{\omega K^{*-}K^+}=g_{\omega K^{*0}\bar{K}^0}=g_{\omega \bar{K}^{*0}K^0}=\frac{1}{\sqrt{2}}g_{VVP},\\
            g_{\rho^0 K^{*+}K^-}&=g_{\rho^0 K^{*-}K^+}=-g_{\rho^0 K^{*0}\bar{K}^0}=-g_{\rho^0 \bar{K}^{*0}K^0}=\frac{1}{\sqrt{2}}g_{VVP},\\
            g_{\omega D^0\bar{D}^{*0}}&=g_{\rho^0 D\bar{D}^{*0}}.
        \end{split}
    \end{align}
    \item $VVV$ vertices
    \begin{align}
        \begin{split}
            g_{\phi K^{*+}K^{*-}}&=-g_{\phi K^{*-}K^{*+}}=g_{\phi K^{*0}\bar{K}^{*0}}=-g_{\phi \bar{K}^{*0}K^{*0}}=g_{VVV},\\
            g_{\omega K^{*+}K^{*-}}&=-g_{\omega K^{*-}K^{*+}}=g_{\omega K^{*0}\bar{K}^{*0}}=-g_{\omega \bar{K}^{*0}K^{*0}}=-\frac{1}{\sqrt{2}}g_{VVV},\\
            g_{\rho^0 K^{*+}K^{*-}}&=-g_{\rho^0 K^{*-}K^{*+}}=-g_{\rho^0 K^{*0}\bar{K}^{*0}}=g_{\rho^0 \bar{K}^{*0}K^{*0}}=-\frac{1}{\sqrt{2}}g_{VVV}.
        \end{split}
    \end{align}
\end{itemize}

\end{appendix}
\bibliographystyle{unsrt}
\bibliography{ref.bib}

\end{document}